\documentclass[aps,prd,reprint,nofootinbib]{revtex4-2}
\usepackage[utf8]{inputenc}
\usepackage{lmodern}
\usepackage[T1]{fontenc}
\usepackage{amsmath}
\usepackage{amsfonts}
\usepackage{mathrsfs}
\usepackage{amssymb}
\usepackage{mathtools}
\usepackage[scr=boondoxo]{mathalpha}
\usepackage{bm}
\usepackage{graphicx}
\usepackage{hyperref}
\hypersetup{
	pdftitle={Second-order gravitational self-force in a highly regular gauge: Covariant and coordinate punctures},
	pdfauthor={Samuel D. Upton},
	pdfdisplaydoctitle,
	colorlinks,
	citecolor=blue
}

\usepackage{physics2}
\usephysicsmodule{ab}
\usephysicsmodule{ab.legacy}
\usephysicsmodule{op.legacy}

\usepackage{xcolor}
\usepackage{cancel}
\usepackage{orcidlink}

\usepackage[italic=true]{derivative}
\newcommand{\dx}{\adif{x}}
\newcommand{\Dx}{\dx}

\renewcommand{\S}{{\rm S}}
\newcommand{\R}{{\rm R}}
\newcommand{\e}{\epsilon}

\DeclareMathOperator*{\sym}{Sym}
\DeclareMathOperator*{\diag}{diag}

\newcommand{\emm}{\mathscr{m}}

\newcommand{\E}{\mathcal{E}}
\newcommand{\B}{\mathcal{B}}
\newcommand{\Edot}{\mathcal{\dot{E}}}
\newcommand{\Bdot}{\mathcal{\dot{B}}}
\newcommand{\nhat}{\hat{n}}
\newcommand{\fullg}{\mathsf{g}}
\newcommand{\DD}{D}
\newcommand{\rb}{\bm{r}}
\newcommand{\rbo}{\rb_0}
\newcommand{\rhob}{\bm{\rho}}
\newcommand{\rhobo}{\rhob_0}
\newcommand{\logsr}{\log\bigl(\tfrac{2 m}{\lambda \rhob}\bigr)}
\newcommand{\Rdot}{\dot{R}}
\newcommand{\acan}{\cancel{\bm{a}}}
\newcommand{\abold}{\bm{a}}

\newcommand{\hdot}{\dot{h}}

\newcommand{\geff}{\tilde{g}}
\newcommand{\Teff}{\tilde{T}}

\newcommand{\BL}{\text{BL}}

\newcommand{\xbar}{{\bar{x}}}
\newcommand{\taubar}{{\bar{\tau}}}
\newcommand{\sbar}{{\bar{\sigma}}}
\newcommand{\ubar}{{\bar{u}}}
\newcommand{\alpbar}{{\bar{\alpha}}}
\newcommand{\betbar}{{\bar{\beta}}}
\newcommand{\mubar}{{\bar{\mu}}}
\newcommand{\nubar}{{\bar{\nu}}}
\newcommand{\gbar}{{\bar{\gamma}}}

\newcommand{\chr}[2]{\Gamma^{#1}_{#2}}

\newcommand{\chruxx}{{\Gamma^{u}_{\Delta\Delta}}}

\newcommand{\chrxxx}{{\Gamma^{\Delta}_{\Delta\Delta}}}
\newcommand{\chrBxx}[1]{\Gamma^{#1}_{\Delta\Delta}}

\newcommand{\chrxBx}[1]{\Gamma^{\Delta}_{#1\Delta}}
\newcommand{\chruBx}[1]{\Gamma^{u}_{#1\Delta}}

\newcommand{\chruxxx}{\Gamma^{u}_{\Delta\Delta,\Delta}}
\newcommand{\chrxxxx}{\Gamma^{\Delta}_{\Delta\Delta,\Delta}}

\newcommand{\calP}{\mathcal{P}}
\newcommand{\calR}{\mathcal{R}}
\newcommand{\deltaG}{\delta G}

\defcitealias{Upton2021}{Paper~I}
\defcitealias{Pound2014}{Paper~II}

\begin{document}

\title{Second-order gravitational self-force in a highly regular gauge: Covariant and coordinate punctures}
\author{Samuel D.\ Upton\,\orcidlink{0000-0003-2965-7674}}
\affiliation{Astronomical Institute of the Czech Academy of Sciences, Bo\v{c}n\'{i} II 1401/1a, CZ-141 00 Prague, Czech Republic}
\affiliation{School of Mathematical Sciences and STAG Research Centre, University of Southampton, Southampton, United Kingdom, SO17 1BJ}
\date{\today}

\begin{abstract}
Gravitational self-force theory is the primary way of modelling extreme-mass-ratio inspirals (EMRIs).
One difficulty that appears in second-order self-force calculations is the strong divergence at the worldline of the small object, which causes both numerical and analytical issues.
Previous work [\href{https://doi.org/10.1103/PhysRevD.95.104056}{Phys. Rev. D 95, 104056 (2017)}; \href{https://doi.org/10.1103/PhysRevD.103.124016}{ibid. 103, 124016 (2021)}] demonstrated that this could be alleviated within a class of highly regular gauges and presented the metric perturbations in these gauges in a local coordinate form.
We build on this previous work by deriving expressions for the highly regular gauge metric perturbations in both fully covariant form and as a generic coordinate expansion.
With the metric perturbations in covariant or generic coordinate form, they can easily be expressed in any convenient coordinate system.
These results can then be used as input into a puncture scheme in order to solve the field equations describing an EMRI.
\end{abstract}

\maketitle

\section{Introduction}\label{sec:intro}

Extreme-mass-ratio inspirals (EMRIs)~\cite{Babak2017} will be a key source of the gravitational waves that will be detected by the Laser Interferometer Space Antenna (LISA), a future space-based gravitational wave detector~\cite{LISA2017,LISAWeb}.
An EMRI features an object of mass \(m \sim 1\text{--}10^2 M_{\odot}\) slowly spiralling into an object of mass \(M \sim 10^{5}\text{--}10^7 M_{\odot}\).
The smaller object is a compact object, such as a black hole or neutron star, whereas the larger object is a supermassive black hole, existing in the centre of most galaxies~\cite{Kormendy1995,Genzel2010,Kormendy2013}.

As the mass ratio, \(\e\coloneqq m/M \sim 10^{-5}\), is very small, the inspiral occurs over a long timescale, with the smaller object expected to complete \(\e^{-1} \sim 10^5\) intricate orbits before plunging into the central black hole~\cite{AmaroSeoane2015,BHroadmap}.
Due to the large number of orbits occurring near to the supermassive black hole, the gravitational waves emitted are expected to provide an excellent picture of the geometry of the black hole in the strong-gravity regime.
This will allow highly accurate tests of general relativity to be performed~\cite{Gair2013,BHroadmap,LISAFundamentalProspects,LISAFundamentalHorizons}.

\subsection{Gravitational self-force}\label{sec:gsf}

The primary method of modelling EMRIs is through a perturbative method known as \emph{gravitational self-force theory}~\cite{Poisson2011,Pound2015Small,Wardell2015,Barack2018,Pound2022}.
The self-force refers to the process by which changes in an external field caused by an object's dynamics propagate back and affect the motion of the very same object.
This method expands the metric describing the geometry of the full spacetime, \(\fullg_{\mu\nu}\), around a known, background metric, \(g_{\mu\nu}\), with perturbations, \(h_{\mu\nu}\), caused by the presence of the small object.
In an EMRI, the disparate sizes of the small and large object lead to a natural perturbative parameter, the mass ratio between the two objects, \(\e \ll 1\).
One can then write the full spacetime metric as the sum of the background spacetime and these perturbations,
\begin{equation}
	\fullg_{\mu\nu} = g_{\mu\nu} + h_{\mu\nu}, \label{eq:fullg}
\end{equation}
where
\begin{equation}
	h_{\mu\nu} = \sum_{n=1}^{\infty}\e^n h^n_{\mu\nu}[\gamma]. \label{eq:hPert}
\end{equation}
In the case of an EMRI, the background metric describes the geometry of the large black hole if it were isolated in spacetime and is taken to be either the Schwarzschild~\cite{Schwarzschild1916,*[Translated in: ]Schwarzschild2003} or Kerr~\cite{Kerr1963} metric.

At the leading order in the mass ratio, the small object's worldline, \(\gamma\), is a geodesic of the background spacetime, \(g_{\mu\nu}\).
The metric perturbations then alter the motion at higher orders and exert a self-force on the body, moving it away from a background geodesic.
This can be written as
\begin{equation}
    \frac{\DD^2 z^\alpha}{\odif{\tau}^2} = \e f^\alpha_1 + \e^2 f^\alpha_2 + \order{\e^3}, \label{eq:geodesicForced}
\end{equation}
which reduces to the geodesic equation when \(\e\to 0\).
In Eq.~\eqref{eq:geodesicForced}, \(z^\alpha\) are coordinates on the accelerated worldline, \(\gamma\), \(\tau\) is the proper time in the background metric, \(g_{\mu\nu}\), \(\DD/\odif{\tau}\coloneqq u^{\mu}\nabla_{\mu}\) is the covariant derivative along the worldline and is compatible with \(g_{\mu\nu}\), \(u^{\alpha}\coloneqq \odv{z^\alpha}/{\tau}\) is the four-velocity and \(f^\alpha_n\) is the \(n\)th-order self-force.
The self-force (or at least part of it) causes the orbit to evolve at a rate of \(\dot{E}/E\sim\e\), resulting in an inspiral over the radiation reaction time, \(t_{rr}\sim E/\dot{E}\sim 1/\e\)~\cite{Pound2015Small}.
Here, \(E\) is the orbital energy and is one of three constants of motion that completely describe the geodesic of a test particle in the background Kerr spacetime; the other two are the azimuthal angular momentum, \(L_z\), and the Carter constant, \(Q\)~\cite{Carter1968}.

One challenge is that we are required to go to second order in the mass ratio in order to model the waveforms accurately.
This is a result of the requirement that for us to extract information from the data gathered by LISA, the phase of the waveform must be accurate to within a fraction of \(1\) radian.
A precise argument for the need for second order was made by \citet{Hinderer2008}.
The orbital parameters, \(J_{B} = \{E,L_{z},Q\}\), slowly evolve over the radiation reaction time, \(t_{rr}\sim 1/\e\).
This motives the introduction of a ``slow time'', \(\tilde{t} = \e t\), so that \(J_{B} = J_{B}(\tilde{t})\).
The orbital frequencies, \(\Omega_{A} = \{\Omega_{r},\Omega_{\theta},\Omega_{\phi}\}\) in the case of Kerr, are functions of the orbital parameters, \(J_{B}(\tilde{t})\), and have perturbative expansions,
\begin{equation}
    \Omega_{A}(J_{B},\e) = \Omega^{(0)}_{A}(J_{B}) + \e\Omega^{(1)}_{A}(J_{B}) + \order{\e^2}, \label{eq:FreqExp}
\end{equation}
where \(\Omega_{A}^{(n\geq 1)}\) are the \(n\)th order corrections to \(\Omega^{(0)}_{A}\) due to the conservative part of the self-force.
The orbital frequencies evolve with respect to the time, \(t\), as
\begin{equation}
    \odv{\Omega_{A}}{t} = \e F^{(1)}_{A}(J_{B}) + \e^{2}F^{(2)}_{A}(J_{B}) + \order{\e^3}. \label{eq:FreqEv}
\end{equation}
where \(F^{(n)}_{A}\) is constructed from the \(n\)th-order dissipative force.
These can then be related to the orbital phases by
\begin{equation}
    \varphi_{A} = \int\Omega_{A}\odif{t}, \label{eq:OrbPhase}
\end{equation}
so that
\begin{equation}
    \varphi_{A} = \frac{1}{\e}\bigl(\varphi^{(0)}_{A}(\tilde{t}) + \e\varphi_{A}^{(1)}(\tilde{t}) + \order*{\e^2}\bigr), \label{eq:PhaseExp}
\end{equation}
where the adiabatic term, \(\varphi^{(0)}_{A}\), is constructed from \(\Omega^{(0)}_{A}\) and \(F^{(1)}_{A}\), and the first post-adiabatic (\(1\)PA) term, \(\varphi^{(1)}_{A}\), is constructed from \(\Omega^{(1)}_{A}\) and \(F^{(2)}_{A}\).
One can see this through noting that an integration over \(t\) introduces a factor of \(1/\e\) through \(\odif{t}=\odv{t}/{\tilde{t}}\odif{\tilde{t}}=\e^{-1}\odif{\tilde{t}}\).
Therefore, to calculate the orbital phases with an error much less than order-\(\e^0\) requires the entirety of the first-order self-force and the dissipative part of the second-order self-force.

It should be stressed that the conservative piece of the first-order self-force and the dissipative piece of the second-order self-force are on equal footing: even if one has the entirety of the first-order self-force (both dissipative and conservative parts), if one does not have the dissipative piece of the second-order self-force then one cannot correctly track the motion of the small object.

As to the current status of the self-force field, at first order, full inspirals driven by the self-force can be computed for generic orbits in the Schwarzschild spacetime for a spinning small object~\cite{Warburton2012,Osburn2016,Warburton2017,vandeMeentWarburton2018}.
One can calculate the full first-order self-force for a non-spinning small object on any generic bound orbit in Kerr~\cite{vandeMeent2018}.
Adiabatic inspirals in Kerr have been performed for equatorial~\cite{Fujita2020} and generic~\cite{Hughes2021} orbits with Ref.~\cite{Lynch2022} performing an equatorial inspiral using the entirety of the first-order self-force.

Second-order calculations are at a much more preliminary stage but important breakthroughs have been made in recent years~\cite{Pound2020,Warburton2021,Durkan2022} with Ref.~\cite{Wardell2023} presenting the first post-adiabatic waveforms for quasicircular orbits in the Schwarzschild spacetime.
Work has also been undertaken on incorporating effects of the spin of the small object as this has an impact at \(1\)PA order on the gravitational-wave phase~\cite{Witzany2019,Piovano2020,Akcay2020,Skoupy2021,Piovano2021,Mathews2022,Skoupy2022,Drummond2022a,Drummond2022b,Skoupy2023,Lynch2023,Witzany2023,Drummond2023}.

\subsection{Local form of the metric perturbations, puncture scheme and infinite mode coupling}\label{sec:background}

\subsubsection{Metric perturbations and effective stress-energy tensor}\label{sec:LocalH}

To find the local form of the metric perturbations, one uses the method of matched asymptotic expansions (for a general introduction to matched asymptotic expansions, see, e.g. Refs.~\cite{Eckhaus1979,Kevorkian1996}, and for an introduction to their use in self-force, see, e.g. Ref.~\cite{Pound2015Small}).
When close to the small object, the expansion from Eqs.~\eqref{eq:fullg}--\eqref{eq:hPert} breaks down as the gravitational field from the small object dominates over that of the background spacetime.
One then introduces a second expansion that focuses in on the small object and then matches this with the external expansion at some appropriate lengthscale.
This is then combined with the vacuum Einstein field equations to solve for the metric perturbations, \(h_{\mu\nu}\).

The metric perturbation can be split into two fields~\cite{Pound2012NLGSF},
\begin{equation}
    h_{\mu\nu} = h^{\R}_{\mu\nu} + h^{\S}_{\mu\nu}, \label{eq:RSSplit}
\end{equation}
where \(h^{\R}_{\mu\nu}\) and \(h^{\S}_{\mu\nu}\) are the \emph{regular field} and \emph{singular field}, respectively.
The regular and singular fields can be expanded in an analogous manner to Eq.~\eqref{eq:hPert}, as
\begin{align}
    h^{\R}_{\mu\nu} ={}& \sum_{n=1}^{\infty} \e^n h^{\R n}_{\mu\nu}, \label{eq:hRExp} \\
    h^{\S}_{\mu\nu} ={}& \sum_{n=1}^{\infty} \e^n h^{\S n}_{\mu\nu}. \label{eq:hSExp}
\end{align}

The regular field has the form of a Taylor series centred on the worldline of the small object and satisfies the vacuum Einstein field equations,
\begin{align}
    \deltaG^{\mu\nu}[h^{\R1}] ={}& 0, \label{eq:hRVacEFE1} \\
    \deltaG^{\mu\nu}[h^{\R2}] ={}& -\delta^{2}G^{\mu\nu}[h^{\R1},h^{\R1}], \label{eq:hRVacEFE2}
\end{align}
throughout the entire spacetime.
When combined with the background metric, it forms a smooth, vacuum effective metric that determines the local geometry that the small object ``feels'',
\begin{equation}
    \geff_{\mu\nu} = g_{\mu\nu} + h^{\R}_{\mu\nu}. \label{eq:geff}
\end{equation}
Through second order, the trajectory of the small object (assuming zero spin) is governed by the equation of motion~\cite{Pound2012,Pound2017}
\begin{multline}
    \frac{\DD^2{z^\mu}}{\odif{\tau^2}} = -\frac{1}{2}(g^{\mu\alpha}+u^{\mu}u^{\alpha})(g_\alpha{}^\delta-h^{\mathrm{R}\delta}_\alpha) \\
        \times (2h^{\mathrm{R}}_{\delta\beta;\gamma} - h^{\mathrm{R}}_{\beta\gamma;\delta}){u^\beta}{u^\gamma} + \order{\epsilon^3}, \label{eq:2ndorderselfforce}
\end{multline}
which can be written as a geodesic in the effective spacetime, \(\geff_{\mu\nu}\), as
\begin{equation}
    \frac{\tilde{\DD}^2{z^\mu}}{\odif{\tilde{\tau}^2}} = \order{\epsilon^3}, \label{eq:genEquivPrinO2}
\end{equation}
where all quantities with tildes are defined with respect to \(\geff_{\mu\nu}\).
This correspondence is known as the generalised equivalence principle~\cite{Pound2017}, which states that (ignoring finite-size effects) a compact object immersed in an external gravitational field will follow a geodesic in some effective metric whose geometry is determined by its own physical mass.

The remaining part of the metric perturbations, the singular field, contains information about the small object's multipole structure~\cite{Pound2012NLGSF}.
Schematically, it has the form
\begin{align}
    h^{\S1}_{\mu\nu} \sim{}& \frac{m}{r}, \label{eq:hS1} \\
    h^{\S2}_{\mu\nu} \sim{}& \frac{m^2 + M^\alpha + S^\alpha}{r^2}, \label{eq:hS2}
\end{align}
where \(r\) is the proper spatial distance to \(\gamma\) and \(M^{\alpha}\)/\(S^{\alpha}\) are the mass/spin dipole terms, respectively.
As in previous work, we enforce that the mass dipole and any higher-order corrections to it vanish.
This ensures that \(\gamma\) tracks the small object's centre of mass~\cite{Gralla2008,Pound2010,Pound2017}.

In certain classes of gauges, the small object also has the effective stress-energy of a point mass in the effective spacetime~\cite{Detweiler2012,Upton2021}.\footnote{This has explicitly been shown in the highly regular gauge and (using a specific distributional definition of the second-order Einstein tensor) in the Lorenz gauge. While it has not been shown, it is likely to hold true in other gauges as well; see the discussion in Sec.~V~E of Ref.~\cite{Upton2021}.}
Using this effective stress-energy tensor, the field equations can be written in the form
\begin{equation}
    \deltaG^{\mu\nu}[\e h^1+\e^{2} h^2] + \e^2\delta^{2}G^{\mu\nu}[h^1,h^1] = 8\pi\Teff^{\mu\nu} + \order{\e^3}, \label{eq:TeffSource}
\end{equation}
where \(\Teff^{\mu\nu}\) is the \emph{Detweiler stress-energy tensor},
\begin{equation}
    \Teff^{\mu\nu} = m\int_{\gamma}\tilde{u}^{\mu}\tilde{u}^{\nu} \frac{\delta^{4}(x-z)}{\sqrt{-\geff}}\odif{\tilde{\tau}}, \label{eq:Teff}
\end{equation}
and all quantities with tildes are defined with respect to the effective metric.
The existence of this stress-energy tensor was first postulated by \textcite{Detweiler2012} and explicitly derived in Ref.~\cite{Upton2021} (hereafter \citetalias{Upton2021}).
One can also write the left-hand side of Eq.~\eqref{eq:TeffSource} in terms of effective quantities as~\cite{Detweiler2012,Upton2021}
\begin{equation}
    \tilde{\deltaG}^{\mu\nu}[h^{\S}] = 8\pi \Teff^{\mu\nu} + \order{\e^3}, \label{eq:dGTeff}
\end{equation}
demonstrating that the system can be described as a linear perturbation of an effective background.

It should be noted that the split into regular and singular fields is not unique~\cite{Pound2014}, but we choose the split to match that of, e.g.\ Refs.~\cite{Pound2012,Pound2012NLGSF,Pound2017,Upton2021}, ensuring that the regular and singular fields satisfy the properties listed above.
That is, the regular field is smooth on the worldline of the small object, forms the effective metric, \(\geff_{\mu\nu}\), and satisfies the generalised equivalence principle.
In addition to the non-uniqueness of the split, it should be emphasised that neither \(h^{\R}_{\mu\nu}\) nor \(h^{\S}_{\mu\nu}\) represent the true physical field; only their sum \(h_{\mu\nu}=h^{\R}_{\mu\nu}+h^{\S}_{\mu\nu}\) does.

We stress that the results discussed in this section are all derived from the principle of matched asymptotic expansions.
One does not start by assuming that the small object is described by a point-particle stress-energy with some effective equation of motion.
Instead, one uses the matching process at each order in \(\e\) to rigorously \emph{derive} these properties from first principles.

\subsubsection{Puncture scheme}\label{sec:punc}

To date, all second-order calculations have involved the use of a \emph{puncture scheme}~\cite{Pound2020,Warburton2021,Wardell2023}; see, e.g.\ Refs.~\cite{Pound2014,Wardell2015,Miller2021,Pound2022} for technical details.\footnote{The first implementations of the puncture scheme were performed at first order in Refs.~\cite{Barack2007a,Barack2007b,Vega2008} although Ref.~\cite{Rosenthal2006} originally suggested its used at second order.}
In this scheme, one introduces a \emph{puncture field}, \(h^{\calP}_{\mu\nu}\approx h^{\S}_{\mu\nu}\), that approximates the singular field to some sufficient order in \(r\) away from the worldline, and goes to zero beyond that.
From this, one can define a \emph{residual field},
\begin{equation}
    h^{\calR}_{\mu\nu}\coloneqq h_{\mu\nu} - h^{\calP}_{\mu\nu}, \label{eq:residualfield}
\end{equation}
so that \(h^{\calR}_{\mu\nu}\approx h^{\R}_{\mu\nu}\) near \(\gamma\).
These fields are then analytically extended down to the worldline, and one solves for the residual field, \(h^{\calR}_{\mu\nu}\), with the puncture field as the source, instead of directly for the physical field, \(h_{\mu\nu}\).

We wish to be able to replace \(h^{\mathrm{R}}_{\mu\nu}\) with \(h^{\mathcal{R}}_{\mu\nu}\) in the equation of motion~\eqref{eq:2ndorderselfforce}.
This is possible if \(h^{\mathcal{R}}_{\mu\nu}\) and its first derivatives are identical to \(h^{\mathrm{R}}_{\mu\nu}\).
To ensure this, we impose the conditions
\begin{gather}
    \lim_{x \to z} \ab(h^{\calP}_{\mu\nu} - h^{\S}_{\mu\nu}) = 0, \label{eq:puncsinglims0th} \\
    \lim_{x \to z} \ab(h^{\calP}_{\mu\nu,\rho} - h^{\S}_{\mu\nu,\rho}) = 0, \label{eq:puncsinglims1st}
\end{gather}
where \(z^\mu\) is a point on the worldline.
Explicitly, to calculate the second-order self-force, we need to go to order \(r\) in our second-order punctures so that our residual field is once differentiable.

Substituting Eq.~\eqref{eq:residualfield} into the field equations and expanding the residual and puncture fields order-by-order, as in Eq.~\eqref{eq:hPert},
\begin{equation}
    h^{\calR/\calP}_{\mu\nu} = \sum_{n=1}^{\infty}\e^{n}h^{\calR/\calP n}_{\mu\nu}, \label{eq:hRPexp}
\end{equation}
we get
\begin{align}
	\deltaG^{\mu\nu}[h^{\calR1}] ={}& -\deltaG^{\mu\nu}[h^{\calP1}], \quad r>0, \label{eq:punc1} \\
	\deltaG^{\mu\nu}[h^{\calR2}] ={}& -\deltaG^{\mu\nu}[h^{\calP2}] - \delta^{2}G^{\mu\nu}[h^1,h^1], \quad r>0. \label{eq:punc2}
\end{align}
These equations can be promoted to the entire domain, including \(r=0\), provided that the puncture field is known to a sufficiently high order in \(r\); see the discussion after Eq.~(13) of \citetalias{Upton2021}.
Combining the field equations with the equation of motion~\eqref{eq:2ndorderselfforce}, one can solve the coupled system of equations and determine how the small object travels in spacetime.

\subsubsection{The problem of infinite mode coupling}\label{sec:infmode}

When implementing the puncture scheme at second order, one encounters the problem of \emph{infinite mode coupling}~\cite{Miller2016}.
To take advantage of the symmetries of the spacetime, one decomposes the metric perturbations into a suitable basis of harmonics.\footnote{Note that the issue described here cannot be avoided by performing a full \(4D\) calculation. Instead of having to go to very high mode numbers in order to obtain convergence of the mode-sum, one would have to perform a very delicate numerical calculation between two terms that diverge as \(1/r^4\).}
For example, in Schwarzschild, one could choose Barack--Lousto--Sago tensor spherical harmonics~\cite{Barack2005,BarackSago2007}, so that the metric perturbations can be decomposed as
\begin{equation}
    h^{n}_{\mu\nu} = \sum_{i\ell\emm}h^n_{i\ell\emm}(t_{\BL},r_{\BL})Y^{i\ell\emm}_{\mu\nu}(\theta,\phi). \label{eq:BLS}
\end{equation}
With the modes written as such, to calculate a single mode of \(\delta^{2}G_{\mu\nu}[h^1,h^1]\) requires one to calculate the infinite sum of products of first-order modes~\cite{Miller2016,SpiersCoupling},
\begin{equation}
    \delta^{2}G_{i\ell\emm}[h^1,h^1] = \sum_{\substack{i_{1}\ell_{1}\emm_{1}\\i_{2}\ell_{2}\emm_{2}}} \mathcal{D}^{i\ell\emm}_{i_{1}\ell_{1}\emm_{1}i_{2}\ell_{2}\emm_{2}}[h^{1}_{i_{1}\ell_{1}\emm_{1}},h^{1}_{i_{2}\ell_{2}\emm_{2}}], \label{eq:d2Gilm}
\end{equation}
where \(\mathcal{D}^{i\ell\emm}_{i_{1}\ell_{1}\emm_{1}i_{2}\ell_{2}\emm_{2}}[h_{i_{1}\ell_{1}\emm_{1}},h_{i_{2}\ell_{2}\emm_{2}}]\) is a certain differential operator~\cite{SpiersCoupling}.
From Eq.~\eqref{eq:hS1}, we see that \(h^{\S1}_{\mu\nu} \sim m/r\).
This means that, generically, the second-order Einstein tensor diverges as \(\sim m^2/r^4\) at the worldline of the small object as it has the structural form, \(\delta^{2}G^{\mu\nu}[h^1,h^1] \sim (\partial h^1)^2 + h^1\partial^{2}h^{1} \sim m^2/r^4\).
After decomposing into modes and integrating over two of the dimensions, one finds that Eq.~\eqref{eq:d2Gilm} acts as
\begin{equation}
    \delta^{2}G_{i\ell\emm}[h^1,h^1] \sim \frac{m^2}{r^2}. \label{eq:d2GilmDiv}
\end{equation}
However, the modes of the first-order field are finite on the worldline~\cite{Barack2009,WardellWarburton2015}, meaning that Eq.~\eqref{eq:d2Gilm} is attempting to reconstruct a divergent function through summing up finite modes.
Thus to get convergence requires one to calculate an arbitrarily large number of modes of the first-order fields to calculate even one second-order mode.

A way to circumvent this problem was provided by \textcite{Miller2016}.
Instead of summing over modes, as in Eq.~\eqref{eq:d2Gilm}, one expands the first-order field into regular and singular pieces.
After expanding the first-order field, the second-order Einstein tensor in the source of the second-order field equations has the form
\begin{multline}
    \delta^{2}G^{\mu\nu}[h^{1},h^{1}] = \delta^{2}G^{\mu\nu}[h^{\R1},h^{\R1}] + 2\delta^{2}G^{\mu\nu}[h^{\R1},h^{\S1}] \\
        + \delta^{2}G^{\mu\nu}[h^{\S1},h^{\S1}], \quad r>0. \label{eq:d2GExp}
\end{multline}
One then replaces the regular and singular fields in Eq.~\eqref{eq:d2GExp} with the residual and puncture fields.
The \(\delta^{2}G_{i\ell\emm}[h^{\calR 1},h^{\calR 1}]\) and \(\delta^{2}G_{i\ell\emm}[h^{\calR 1},h^{\calP 1}]\) terms are sufficiently well-behaved that one may compute the modes directly from the modes of the first-order residual and puncture fields.
As described in Ref.~\cite{Miller2016}, the problem is entirely caused by the slow converge of the modes of \(\delta^{2}G_{i\ell\emm}[h^{\calP 1},h^{\calP 1}]\) as this is the term that causes the non-mode-decomposed second-order Einstein tensor to diverge as \(\sim m^2/r^4\).
Instead of summing up the products of the modes of \(h^{\calP 1}_{\mu\nu}\), \textcite{Miller2016} directly calculate \(\delta^{2}G^{\mu\nu}[h^{\calP 1},h^{\calP 1}]\) in four dimensions using the four dimensional expression for \(h^{\calP 1}_{\mu\nu}\) and then decompose this quantity into modes.
Unfortunately, while this makes the calculation of the modes of the source possible, it is incredibly computationally expensive and takes up almost all the code runtime when implemented (such as in Ref.~\cite{Pound2020}).
This is due to having to calculate the modes by numerically integrating the complete four-dimensional expression on a grid of \(r_{\BL}\) and \(r\) values.
This will not be efficiently extendible when approaching problems involving more complicated dynamics, such as generic orbits in Kerr.

\subsection{Highly regular gauge}\label{sec:hr}

The highly regular gauge was introduced by \textcite{Pound2017} to ameliorate the strong divergences that occur near the worldline of the small object when in a generic gauge.
In this gauge, the most singular piece of the second-order perturbation now has the form \(\sim m^{2}r^{0}\) instead of the \(\sim m^2/r^2\) behaviour previously seen; see Refs.~\cite{Pound2017,Upton2021} for a full discussion.
One can divide the second-order singular field into two pieces: a ``singular times regular'' piece, \(h^{\S\R}_{\mu\nu}\sim m h^{\R1}_{\mu\nu}/r\), and a ``singular times singular'' piece, \(h^{\S\S}_{\mu\nu} \sim m^{2}r^0\).
By simple order counting of \(m\) and \(h^{\R1}_{\mu\nu}\), we see that, in the second-order Einstein field equations, \(h^{\S\S}_{\mu\nu}\) is sourced by \(\delta^{2}G^{\mu\nu}[h^{\S1},h^{\S1}]\), as they both feature terms \(\sim m^2\), and that \(h^{\S\R}_{\mu\nu}\) is sourced by \(\delta^{2}G^{\mu\nu}[h^{\R1},h^{\S1}]\) as both expressions have terms of the form \(\sim m h^{\R1}_{\mu\nu}\).
Although the \(h^{\S\R}_{\mu\nu}\) term appears more divergent, as discussed in \citetalias{Upton2021}, its source, \(\delta^{2}G^{\mu\nu}[h^{\R1},h^{\S1}]\), is well defined as a distribution.
The ``singular times singular'' term causes the most issues.
Acting on the ``singular times singular'' piece with the linearised Einstein operator, we see that \(\deltaG^{\mu\nu}[h^{\S\S}] \sim m^{2}/r^2\).
Therefore, we know that the most singular piece of the second-order Einstein tensor can only act as badly \(\delta^{2}G^{\mu\nu}[h^{\S1},h^{\S1}] \sim m^2/r^2\) instead of \(\sim m^2/r^4\) as in a generic gauge.
This means that when decomposing into modes, the individual modes of the second-order Einstein tensor can behave, at worst, as \(\delta^{2}G_{i\ell\emm}[h^1,h^1]\sim m^{2}\log|r|\).
While this is still divergent, it is much weaker than in the Lorenz gauge.

The highly regular gauge enforces that the local light cone structure around \(\gamma\) is preserved in the perturbed spacetime.
To do so, two gauge conditions are imposed on the singular field.
Firstly, the metric perturbations vanish when contracted with \(k^\mu\), the null vector tangent to the future light cone that emanates from the worldline:
\begin{equation}
    h^{\S}_{\mu\nu}k^{\mu} = 0. \label{eq:HRGaugeCondk}
\end{equation}
Secondly, the perturbations are trace-free with respect to \(\Omega_{AB}\), the metric on surfaces of constant luminosity distance:
\begin{equation}
    h^{\S}_{\mu\nu}e^{\mu}_{A}e^{\nu}_{B}\Omega^{AB} = 0, \label{eq:HRGaugeTrace}
\end{equation}
where an upper case Latin letter indicates a quantity defined on those surfaces and \(e^\mu_A \coloneqq \partial x^\mu/\partial\theta^A\) is the basis vector, where \(x^\mu\) are coordinates in the full spacetime and \(\theta^A\) are coordinates on the surface of constant luminosity distance.
These gauge conditions ensure that the local background light cone structure is preserved in the perturbed spacetime and that the background luminosity distance is an affine parameter on the null rays that generate the light cones.
\begin{figure}
    \centering
    \includegraphics[width=0.9\columnwidth]{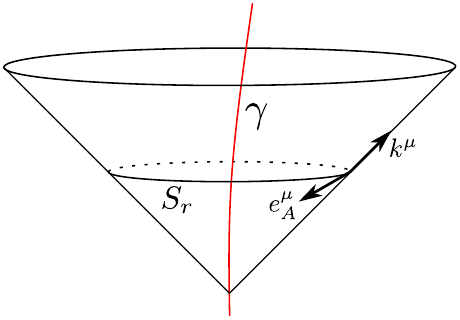}
    \caption[Gauge conditions for the highly regular gauge]{Geometric picture of the gauge conditions for the highly regular gauge. The image features a light cone emanating from the worldline, \(\gamma\). The null vector, \(k^\mu\), is tangent to the light cone along radially outgoing curves, and the basis vector, \(e^\mu_A\), is tangent to the light cone along spheres of constant luminosity distance, \(S_r\). Based on Fig.~16 from Ref.~\cite{PoundAGR}.}
    \label{fig:HRGaugeConditions}
\end{figure}
An image showing the geometric construction is given in Fig.~\ref{fig:HRGaugeConditions}.

When working with a puncture scheme, one can impose different gauge conditions on the residual and puncture fields; see the discussions in Sec.~IV~A of Ref.~\cite{Gralla2012}, Sec.~VII~A of Ref.~\cite{Pound2017} and Sec.~VI~A of \citetalias{Upton2021}.
Therefore, to control the singularity structure, one can impose the highly regular gauge conditions on the puncture.
Then, one can impose any convenient gauge conditions on the residual field that simplify the left-hand side of the field equations~\eqref{eq:punc1}--\eqref{eq:punc2}.

Reference~\cite{Pound2017} only provided the leading-order pieces of the second-order metric perturbations in this gauge.
\citetalias{Upton2021} extended this to include all orders needed to perform a numerical calculation of the self-force.
These expressions were provided in Fermi--Walker coordinates, a particular coordinate system that is tethered to an accelerated worldline, \(\gamma\), and is useful for analysing the properties of fields near to this worldline.
However, in order to use the expressions in a puncture scheme, one needs to write them in a coordinate scheme specialised to the problem at hand, such as Boyer--Lindquist coordinates \((t_{\BL},r_{\BL},\theta_{\BL},\phi_{\BL})\)~\cite{Boyer1967}.
To avoid a potentially complicated coordinate transformation from Fermi--Walker coordinates to the new coordinate system, one can convert the Fermi--Walker expressions into covariant form.
This can then be written in the chosen coordinate system.

To do so, one can use the method given by \textcite{Pound2014} (hereafter \citetalias{Pound2014}).
This method was developed in order to transform expressions for the singular field in the Lorenz gauge into covariant form.
These expressions, after being written in an appropriate coordinate system and decomposed into a suitable basis of modes, were used as input into the two-timescale expansion~\cite{Miller2021} that has been used in the only existing calculations of second-order quantities~\cite{Pound2020,Warburton2021,Wardell2023}.

The aim in \citetalias{Pound2014} was the same as the aim here: to convert expressions for the singular field written in Fermi--Walker coordinates into fully covariant expressions.
This covariant expression can then be used as input into the previously mentioned puncture scheme.

\subsection{Paper outline}\label{sec:outline}

We begin in Secs.~\ref{sec:LocalExp} and ~\ref{sec:FWtoCov} by recapping local expansion methods using bitensors; tensorial functions of two spacetime points; the construction of Fermi--Walker coordinates, and the conversion from Fermi--Walker coordinates to covariant form, as introduced by \citetalias{Pound2014}.
Readers familiar with these concepts should feel free to skip directly to Sec.~\ref{sec:HRCov}, where the covariant punctures for the metric perturbations in the highly regular gauge are derived.
These are displayed in an abridged form due to their length, but the full expressions are provided in a \textsc{Mathematica} notebook in the Supplemental Material~\cite{SuppMat}.

Section~\ref{sec:CoordExp} then re-expands the covariant expressions from Sec.~\ref{sec:HRCovPuncFinal} into a generic coordinate expansion.
The method for re-expanding the various covariant quantities is detailed in Sec.~\ref{sec:bitensorExp} and, as before, readers familiar with this method can skip directly to Sec.~\ref{sec:CoordPunc} where the generic coordinate expansions are presented.
As with the covariant expressions, the coordinate punctures are too lengthy to include fully in this paper and are provided in the Supplemental Material~\cite{SuppMat}.

Finally, we sum up the findings of this paper in Sec.~\ref{sec:conclusion} and discuss potential future avenues for research.

\subsection{Conventions and definitions}\label{eq:defs}

We use metric signature \((-,+,+,+)\) and geometric units with \(c=G=1\).
Indices using Greek letters run from \(0\) to \(3\) and with lowercase Latin letters run from \(1\) to \(3\).
Greek/Latin indices are raised and lowered from the background metric, \(g_{\mu\nu}\), and the flat-space Euclidean metric, \(\delta_{ab}\), respectively.

A primed index on a tensor, \(A^{\mu'}\), indicates the tensor is evaluated at \(x'^\mu\coloneq z^\mu(\tau)\), where \(z^\mu(\tau)\) are coordinates on the worldline, \(\gamma\).
An unprimed index on a tensor, \(A^\mu\), is used for when the tensor is evaluated away from the worldline at \(x^\mu\).
An overset bar on a tensorial index, \(A^{\mubar}\), is used when a tensor is evaluated at \(\xbar^{\mu}\).
This is a point on the worldline which is connected to \(x^\mu\) by an orthogonal geodesic.

A hat on a tensor, \(\hat{T}^{a_{1}\ldots a_{i}}\), refers to the symmetric trace-free (STF) part of the tensor with respect to the flat-space metric, \(\delta_{ab}\).
The covariant derivative is given by \(\nabla\) or a semi-colon and is compatible with the background metric, \(g_{\mu\nu}\).
The partial derivative is given by \(\partial\) or a comma.

We adopt notation from Ref.~\cite{Haas2006} for contractions of \(u^{\mu'}\), \(\sigma^{\mu'}\) and \(\Dx^{\mu'}\) so that,
\begin{align}
    \Gamma^{\Delta}_{u\Delta,\Delta} \coloneqq \Gamma^{\alpha'}_{\beta'\mu',\nu'}\Dx_{\alpha'}u^{\beta'}\Dx^{\mu'}\Dx^{\nu'}, \\
    \Rdot_{u\sigma u\sigma} \coloneqq R_{\alpha'\beta'\mu'\nu';\gamma'}\sigma^{\beta'}\sigma^{\nu'}u^{\alpha'}u^{\mu'}u^{\gamma'},
\end{align}
for example.
We use analogous notation for contractions of tensors evaluated at \(\xbar^{\mu}\), e.g.
\begin{align}
    \Rdot_{\ubar\sbar\ubar\sbar} \coloneqq R_{\alpbar\betbar\mubar\nubar;\gbar}\sigma^{\betbar}\sigma^{\nubar}u^{\alpbar}u^{\mubar}u^{\gbar}.
\end{align}

The calculations in this paper make extensive use of \textsc{Wolfram Mathematica}~\cite{Mathematica} and the tensor algebra package \textsc{xAct}~\cite{xAct,xPerm,xPert,xTras,SymManipulator,TexAct}.

\section{Local expansion methods}\label{sec:LocalExp}

In this section, we recap the methods of performing covariant and coordinate expansions of tensorial quantities near the worldline.
We also give an overview of the construction of Fermi--Walker coordinates.

\subsection{Covariant expansions using bitensors}\label{sec:CovExp}

In this section, we outline how one may construct local covariant expansions of tensor fields.
Our explanation of the method follows that of Refs.~\cite{Synge1960,Christensen1976,Poisson2011}.
To do this, we introduce the concept of a bitensor: a tensor which is a function of two spacetime points.
One important bitensor that we will make extensive use of is Synge's world function~\cite{Synge1960,Poisson2011},
\begin{equation}
	\sigma(x,x') = \frac{\varepsilon}{2}\Bigl(\int_{\beta}\odif{s}\Bigr)^2, \label{eq:syngeDef}
\end{equation}
where \(\beta\) is the unique geodesic connecting \(x^\mu\) and \(x^{\mu'}\), \(s\) is an affine parameter and \(\varepsilon = \mp 1\) for time/spacelike geodesics (not to be confused with the mass ratio \(\e\)).
This gives half the geodesic distance squared between the points \(x^\mu\) and \(x^{\mu'}\).
If the two points are connected by a null geodesic, then \(\sigma(x,x')\) is identically zero.
We will use \(\lambda\) as a formal order counting parameter to count powers of spatial distance away from the worldline, \(\gamma\), so that \(\sigma\sim\lambda^2\).

We denote derivatives of Synge's world function as \(\sigma_{\mu'} \coloneqq \nabla_{\mu'}\sigma(x,x') = \partial_{\mu'}\sigma(x,x')\).
Note also that we may take derivatives of Synge's world function at the unprimed coordinates as well, giving
\(\sigma_{\mu} \coloneqq \nabla_{\mu}\sigma(x,x') = \partial_{\mu}\sigma(x,x')\).
This can be generalised to higher and higher derivatives, e.g. \(\sigma_{\mu'\nu'} \coloneqq \nabla_{\nu'}\nabla_{\mu'}\sigma\) or \(\sigma_{\mu'\nu} \coloneqq \nabla_{\nu}\nabla_{\mu'}\sigma\).
The indices of \(\sigma\) tell us its tensorial structure at both \(x^\mu\) and \(x^{\mu'}\), that is, \(\sigma_{\mu'\nu'}\) is a rank-\(2\) tensor at \(x^{\mu'}\) but a scalar at \(x^\mu\).
Likewise, \(\sigma_{\mu'\nu}\) is a covector at both \(x^\mu\) and \(x^{\mu'}\).
This property demonstrates that we can always commute primed and unprimed indices as the existence of one does not affect the tensorial rank at the other point.
Derivatives of Synge's world function also satisfy the useful identity
\begin{equation}
	g_{\alpha\beta}\sigma^{\alpha}\sigma^{\beta} = g_{\alpha'\beta'}\sigma^{\alpha'}\sigma^{\beta'} = 2\sigma(x,x'). \label{eq:syngeIdentity}
\end{equation}
By taking derivatives of Eq.~\eqref{eq:syngeIdentity} and then the limit as \(x^\mu\) goes to \(x^{\mu'}\), one may derive local covariant expansions of \(\sigma_{\alpha'\ldots\alpha\ldots}\) in terms of quantities defined on the worldline.
To see an example, we start by introducing the standard notation for the coincidence limit~\cite{Synge1960},
\begin{equation}
	[A^{\alpha\ldots\alpha'\ldots}_{\beta\ldots\beta'\ldots}] \coloneqq \lim_{x^{\mu}\to x^{\mu'}} A^{\alpha\ldots\alpha'\ldots}_{\beta\ldots\beta'\ldots}(x,x'). \label{eq:CoinLim}
\end{equation}
It immediately follows from Eqs.~\eqref{eq:syngeDef}--\eqref{eq:syngeIdentity} that
\begin{equation}
	[\sigma] = [\sigma_{\alpha}] = [\sigma_{\alpha'}] = 0, \label{eq:SigCoin}
\end{equation}
as, if the length of \(\beta\) goes to \(0\), then the integral in Eq.~\eqref{eq:syngeDef} vanishes.
Taking primed derivatives of Eq.~\eqref{eq:syngeIdentity}, we see
\begin{equation}
	\sigma_{\mu'} = \sigma^{\nu'}\sigma_{\nu'\mu'}, \label{eq:sigmaBBEq}
\end{equation}
which implies that
\begin{equation}
	[\sigma_{\mu'\nu'}] = g_{\mu'\nu'}. \label{eq:sigmaBBCoin}
\end{equation}
This can be repeated to find higher and higher derivatives of \(\sigma(x,x')\)~\cite{Christensen1976},
\begin{align}
	[\sigma_{\mu'\nu'\rho'}] ={}& 0, \label{eq:sigmaBBBCoin} \\
	[\sigma_{\mu'\nu'\alpha'\beta'}] ={}& \frac{2}{3}R_{\mu'(\alpha'\beta')\nu'}. \label{eq:sigmaBBBBCoin}
\end{align}

Another object we will make use of is the parallel propagator, \(g^{\mu'}{}_{\mu}(x,x')\)~\cite{Synge1960,Christensen1976,Poisson2011}.
The parallel propagator parallel transports a tensor from \(x^{\mu'}\) to \(x^\mu\) along \(\beta\).
For instance, the vector \(A^\mu(x)\) can be transported from/to \(A^{\mu'}(x')\) via
\begin{align}
	A^{\mu}(x) ={}& g^{\mu}{}_{\mu'}(x,x')A^{\mu'}(x'), \label{eq:PrimeToA} \\
	A^{\mu'}(x') ={}& g^{\mu'}{}_{\mu}(x',x)A^{\mu}(x), \label{eq:AToPrime}
\end{align}
respectively.
These expressions hold for covectors as well and tensors with any number of indices with the inclusion of an appropriate number of parallel propagators, e.g.
\begin{equation}
	A^{\alpha\beta}{}_{\mu}{}^{\nu}(x) = g^{\alpha}{}_{\alpha'}g^{\beta}{}_{\beta'}g^{\mu'}{}_{\mu}g^{\nu}{}_{\nu'}A^{\alpha'\beta'}{}_{\mu'}{}^{\nu'}(x'). \label{eq:PrimeToAMoreIndices}
\end{equation}
It also has the properties that when contracted with itself, it returns the Kronecker delta,
\begin{align}
	g^{\mu}{}_{\mu'}g^{\mu'}{}_{\nu} ={}& \delta^\mu_\nu, \label{eq:ppDelta} \\
	g^{\mu'}{}_{\mu}g^{\mu}{}_{\nu'} ={}& \delta^{\mu'}_{\nu'}, \label{eq:ppDeltaPrime}
\end{align}
and is symmetric in indices and arguments,
\begin{equation}
	g_{\mu}{}^{\mu'}(x,x') = g^{\mu'}{}_{\mu}(x',x). \label{eq:ppSym}
\end{equation}
When contracted with Synge's world function, it gives
\begin{align}
	\sigma_{\mu} ={}& -g^{\mu'}{}_{\mu}\sigma_{\mu'}, \label{eq:SigXPToX} \\
	\sigma_{\mu'} ={}& -g^{\mu}{}_{\mu'}\sigma_{\mu}, \label{eq:SigXToXP}
\end{align}
and its derivative contracted with Synge's world function vanishes for all combinations of primed and unprimed indices, e.g.\
\begin{equation}
	g^{\mu'}{}_{\mu;\nu}\sigma^{\nu} = 0. \label{eq:SigPPDContract}
\end{equation}
As we did for Synge's world function with Eq.~\eqref{eq:syngeIdentity}, we can calculate different covariant expansions by repeatedly differentiating Eq.~\eqref{eq:SigPPDContract} and taking the coincidence limit.
For example~\cite{Christensen1976},
\begin{align}
	[g^{\mu}{}_{\nu'}] ={}& \delta^{\mu'}_{\nu'}, \label{eq:PPCoin} \\
	[g^{\mu}{}_{\nu';\alpha'}] ={}& 0, \label{eq:PPDCoin} \\
	[g^{\mu}{}_{\nu';\alpha\beta}] ={}& -\frac{1}{2}R^{\mu'}{}_{\nu'\alpha'\beta'}. \label{eq:PPDDCoin}
\end{align}

Combining the previous definitions, we can then express an arbitrary tensor \(A^{\mu}{}_{\nu}\), evaluated at \(x\), in terms of quantities evaluated at \(x'\) as
\begin{multline}
	A^{\mu}{}_{\nu}(x) = g^{\mu}{}_{\mu'}g_{\nu}{}^{\nu'}\Bigl(A^{(0)\mu'}{}_{\nu'}(x') + \lambda A^{(1)\mu'}{}_{\nu'\alpha'}(x')\sigma^{\alpha'} \\
	+ \frac{\lambda^2}{2}A^{(2)\mu'}{}_{\nu'\alpha'\beta'}(x')\sigma^{\alpha'}\sigma^{\beta'}\Bigr) + \order{\lambda^3}, \label{eq:xxprimeexp}
\end{multline}
where \(\lambda\) is a formal order counting parameter to be set to unity at the end of the calculation.
The unknown coefficients, \(A^{(N)\mu'}{}_{\nu'{\alpha'}_{1}\ldots{\alpha'}_{n}}\), can be found in the same manner as before by repeated differentiation and taking of the coincidence limit.
As an example, we seek the covariant expansion of \(\sigma_{\mu'\nu'}\).
We first expand, as in Eq.~\eqref{eq:xxprimeexp} but without the need for parallel propagators, as
\begin{align}
	\sigma_{\mu'\nu'} ={}& \sigma^{(0)}_{\mu'\nu'} + \lambda\sigma^{(1)}_{\mu'\nu'\alpha'}\sigma^{\alpha'} + \frac{\lambda^2}{2}\sigma^{(2)}_{\mu'\nu'\alpha'\beta'}\sigma^{\alpha'}\sigma^{\beta'} \nonumber \\
		& + \order{\lambda^3}. \label{eq:sigmaCovExp}
\end{align}
We know from Eq.~\eqref{eq:sigmaBBCoin}, that \(A^{(0)}_{\mu'\nu'} = g_{\mu'\nu'}\).
Taking primed derivatives and the coincidence limit gives that
\begin{align}
	\sigma^{(1)}_{\mu'\nu'\alpha'} ={}& [\sigma_{\mu'\nu'\alpha'}] = 0, \label{eq:sig1CovExp} \\
	\sigma^{(2)}_{\mu'\nu'\alpha'\beta'} ={}& [\sigma_{\mu'\nu'\alpha'\beta'}] = \frac{2}{3}R_{\mu'(\alpha'\beta')\nu'}, \label{eq:sig2CovExp}
\end{align}
meaning that
\begin{equation}
	\sigma_{\mu'\nu'} = g_{\mu'\nu'} + \frac{\lambda^2}{3}R_{\mu'\alpha'\beta'\nu'}\sigma^{\alpha'}\sigma^{\beta'} + \order{\lambda^3}. \label{eq:sigmaCovExpFinal}
\end{equation}
This can be repeated for any required covariant quantity.
Ref.~\cite{Ottewill2011} provides a semi-recursive method for calculating expansions of Synge's world function and the parallel propagator, along with many other covariant quantities.

\subsection{Fermi--Walker coordinates}\label{sec:FWcoords}

To analyse the properties of the fields near the worldline of the small object, we introduce Fermi--Walker coordinates, \((t,x^a)\), attached to the accelerated worldline, \(\gamma\).
Our description of Fermi--Walker coordinates summarises that of Refs.~\cite{Poisson2004,Poisson2011}.
To begin, we introduce an orthonormal tetrad, \((u^{\mu},e^{\mu}_{a})\), on \(\gamma\) which is defined at the point \(z(\tau)\) so that it satisfies
\begin{align}
	\frac{\DD e^\mu_a}{\odif{\tau}} ={}& a_{\nu}e^\nu_{a}u^{\mu}, \label{eq:FWTetradTransport} \\
	g_{\mu\nu}u^{\mu}u^{\nu} ={}& -1, \label{eq:FWUNorm} \\
	g_{\mu\nu}e^\mu_{a}u^\nu ={}& 0, \label{eq:FWUSpatialNorm} \\
	g_{\mu\nu}e^\mu_{a}e^\nu_{b} ={}& \delta_{ab}, \label{eq:FWSpatialNorm}
\end{align}
where \(u^\mu = \odv{z^\mu}/{\tau}\) is the curve's four-velocity, \(a^\mu = \DD^2z^\mu/\odif{\tau}^2\) is the acceleration of \(\gamma\) and \(\delta_{ab}=\diag(1,1,1)\) is the three-dimensional flat space metric.
If \(\gamma\) is a geodesic then \(a^\mu\) vanishes.
Equation~\eqref{eq:FWTetradTransport} ensures that the tetrad basis is Fermi--Walker transported along \(\gamma\), thus keeping it orthogonal to the worldline as it travels along it.
This condition reduces to that of parallel transport when the worldline is a geodesic.
Equations~\eqref{eq:FWUNorm}--\eqref{eq:FWSpatialNorm} then ensure that it is orthonormal at all points on \(\gamma\).
The dual tetrad, \((e^0_\mu,e^a_\mu)\), can be defined as satisfying
\begin{align}
	e^0_\mu ={}& -u_\mu, \label{eq:FWTetradInvU} \\
	e^a_\mu ={}& \delta^{ab}g_{\mu\nu}e^{\nu}_{b}. \label{eq:FWTetradInvA}
\end{align}
Equations~\eqref{eq:FWUNorm}--\eqref{eq:FWTetradInvA} then imply that we can write the metric and inverse metric as
\begin{align}
	g_{\mu\nu} ={}& -e^0_{\mu}e^0_{\nu} + \delta_{ab}e^{a}_{\mu}e^{b}_{\nu}, \label{eq:FWgRel} \\
	g^{\mu\nu} ={}& -u^{\mu}u^{\nu} + \delta^{ab}e_{a}^{\mu}e_{b}^{\nu}, \label{eq:FWgInvRel}
\end{align}
respectively.

With the orthonormal tetrad constructed, we may now create a local coordinate system so that we may derive the form of the metric near \(\gamma\).
The full technical details are not considered here (see Ref.~\cite[Chs.~9.3--9.5]{Poisson2011} for more details) but we outline the geometric picture of the coordinate construction.
At a point \(\xbar \coloneqq z(t)\) on \(\gamma\), where \(t\) is the proper time, we generate a surface orthogonal to the worldline by emitting spacelike geodesics from \(z(t)\) that are orthogonal to \(\gamma\).
We can then label a point on this surface with coordinates \(x^a\) so that we have coordinates, \((t,x^a)\), that describe points near to the worldline.
The tetrad can be written in terms of Synge's world function as
\begin{gather}
	x^0 = t, \label{eq:DFNCX0} \\
	x^a = -e^{a}_{\alpbar}(\xbar)\sigma^{\alpbar}(x,\xbar), \label{eq:FNCXA} \\
	\sigma_{\alpbar}(x,\xbar)u^{\alpbar}(\xbar) = 0. \label{eq:SigUOrth}
\end{gather}
As stated previously, Synge's world function gives half the geodesic distance squared between two points (up to a minus sign) meaning that a derivative gives the geodesic distance.
This quantity is then contracted with the spatial Fermi--Walker tetrad leg, \(e^a_{\alpbar}\), to give the Fermi--Walker spatial distance, \(x^a\).
The third equation ensures that \(\sigma_{\alpbar}\) is always orthogonal to the worldline.
Alternatively, we can write \(x^i = rn^i\), with \(r \coloneqq \sqrt{\delta_{ab}x^{a}x^{b}} = \sqrt{2\sigma(x,\xbar)}\) being the proper distance (along a unique spacelike geodesic orthogonal to \(\gamma\)) from \(\gamma\) to the point being considered and \(n^i\) being a unit vector giving the direction that the point lies in respective to \(\gamma\).
We note as well that, as with \(\sigma_{\alpha'}\), \(r\sim\lambda\) and so counts powers of distance from the worldline.
\begin{figure}
	\centering
	\includegraphics[width=0.7\columnwidth]{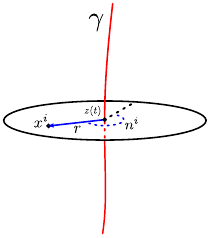}
	\caption[Geometric representation of Fermi--Walker coordinates]{Visualisation of construction of Fermi--Walker coordinates. At the point \(z(t)\), we generate an orthogonal surface and label points on that surface with the coordinate \(x^i\). The quantity \(r\) gives the proper distance to \(x^i\) and \(n^i\) picks out the unique orthogonal geodesic that connects \(x^i\) and \(\gamma\). Based on Fig.~6 from Ref.~\cite{Poisson2011}.}
	\label{fig:FNC}
\end{figure}
A geometric representation of the Fermi--Walker coordinate construction is given in Fig.~\ref{fig:FNC}.

Using these coordinates, we can write the metric near \(\gamma\) in the form~\cite{Pound2017}
\begin{subequations}
	\label{eq:FWmetric}
	\begin{align}
		g_{tt} ={}& -1 - 2ra_{i}n^{i} - r^2(R_{titj} + a_{i}a_{j})n^{ij} \nonumber \\
		& - \frac{r^3}{3}(4R_{titj}a_{k} + R_{titj;k})n^{ijk} + \order{r^4}, \label{eq:FWmetrictt} \\
		g_{ta} ={}& -\frac{2r^2}{3}R_{tiaj}n^{ij} - \frac{r^3}{12}(4R_{tiaj}a_k + 3R_{tiaj;k})n^{ijk} \nonumber \\
		& + \order{r^4}, \label{eq:FWmetricta} \\
		g_{ab} ={}& \delta_{ab} - \frac{r^2}{3}R_{aibj}n^{ij} - \frac{r^3}{6}R_{aibj;k}n^{ijk} + \order{r^4}, \label{eq:FWmetricab}
	\end{align}
\end{subequations}
where all Riemann terms are evaluated on \(\gamma\) at time \(t\).
When evaluating Eq.~\eqref{eq:FWmetric} on \(\gamma\), we immediately see that the metric in Fermi--Walker coordinates reduces to the Minkowski metric.
However, the Christoffel symbols at lowest order are not all zero.
Instead, \(\chr{t}{ta}|_{\gamma} = a_a\) and \(\chr{a}{tt}|_{\gamma} = a^a\); both reduce to \(0\) if \(\gamma\) is a geodesic.

As we are looking at a vacuum solution with \(R_{\mu\nu} = 0\), we may use the identities from Appendix~D3 of Ref.~\cite{Poisson2010} to write
\begin{subequations}
	\label{eq:R_tidal}
	\begin{align}
		R_{tatb} ={}& \E_{ab}, \label{eq:R2t_tidal} \\
		R_{abct} ={}& \e_{ab}{}^i\B_{ic}, \label{eq:R1t_tidal} \\
		R_{abcd} ={}& -\epsilon_{abi}\epsilon_{cdj}\E^{ij} \label{eq:R0t_tidal}
	\end{align}
\end{subequations}
and the derivatives as
\begin{subequations}
	\label{eq:R_deriv_tidal}
	\begin{align}
		R_{tatb;c} ={}& \E_{abc} + \frac{2}{3}\e_{ci(a}\Bdot_{b)}{}^i, \label{eq:R2t_deriv_tidal} \\
		R_{abct;d} ={}& \e_{ab}{}^i\Bigl(\frac{4}{3}\B_{icd} - \frac{2}{3}\e_{dj(i}\Edot^{j}{}_{c)}\Bigr), \label{eq:R1t_deriv_tidal} \\
		R_{abcd;e} ={}& -\epsilon_{abi}\epsilon_{cdj}\Bigl(\E^{ij}{}_e + \frac{2}{3}\e_{ek}{}^{(i}\Bdot^{j)k}\Bigr). \label{eq:R0t_deriv_tidal}
	\end{align}
\end{subequations}
The quantities \(\E\) and \(\B\) are the tidal moments felt by an extended body moving on the world line, \(\gamma\), where two/three indices refer to the quadrupole/octopole moments respectively.
They are symmetric and trace-free, with respect to \(\delta_{ab}\), over all indices and only depend on the proper time, \(t\).

\section{Converting Fermi--Walker coordinates to covariant form}\label{sec:FWtoCov}

In this section we review the method used in \citetalias{Pound2014} to derive the covariant Lorenz gauge puncture.
While the full technical details containing derivations of the various quantities are contained within that paper, we reproduce the essential results that we will need to produce the highly regular gauge puncture.
The final results will be covariant quantities expressed entirely in terms of parallel propagators, the four-velocity, Riemann tensors, and Synge's world function.

The idea behind the method from \citetalias{Pound2014} is to express the field at a point \(x\) in terms of an arbitrary nearby point on the worldline, \(x' = z(\tau')\).
This is done through an intermediary point, \(\xbar = z(\taubar)\), which lies on \(\gamma\) and is separated from \(x'\) by the difference in proper time
\begin{equation}
	\Delta\tau \coloneqq \taubar - \tau'. \label{eq:deltatau}
\end{equation}
The intermediary point, \(\xbar\), is then connected to \(x\) by the unique geodesic that intersects the worldline orthogonally.
\begin{figure}
	\centering
	\includegraphics[width=0.6\columnwidth]{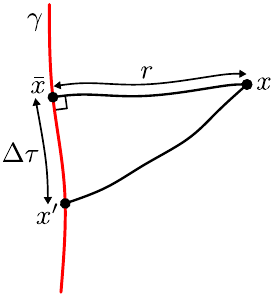}
	\caption[Relationship between \(x\), \(x'\) and \(\xbar\).]{Diagram illustrating the relationship between \(x\), \(x'\) and \(\xbar\). The two points \(x'\) and \(\xbar\) are points on the worldline, \(\gamma\), separated by \(\Delta\tau\) while \(\xbar\) and \(x\) are connected by the geodesic that intersects \(\gamma\) orthogonally. Based on Fig.~1 from \citetalias{Pound2014}.}
	\label{fig:xRel}
\end{figure}
A visual representation is provided in Fig.~\ref{fig:xRel}.

As Fermi--Walker coordinates are constructed geometrically, see Sec.~\ref{sec:FWcoords}, there is a very straightforward way to convert them into covariant form.
We know from Eqs.~\eqref{eq:DFNCX0}--\eqref{eq:SigUOrth}, that there is a simple correspondence between Fermi--Walker coordinates and covariant quantities.
As we saw in the text below Eq.~\eqref{eq:SigUOrth}, we can write the Fermi--Walker radial distance in terms of covariant quantities with
\begin{equation}
	r \coloneqq \sqrt{\delta_{ab}x^{a}x^{b}} = \sqrt{P_{\alpbar\betbar}\sigma^{\alpbar}\sigma^{\betbar}} = \sqrt{2\sbar}, \label{eq:rsig}
\end{equation}
where
\begin{equation}
	\sbar \coloneqq \sigma(x,\xbar). \label{eq:sbar}
\end{equation}
We have added an extra step in Eq.~\eqref{eq:rsig}, where we have rewritten the flat-space metric in terms of the projection operator,
\begin{equation}
	e^\alpha_a e^{a\beta} = P^{\alpha\beta} = g^{\alpha\beta} + u^{\alpha}u^{\beta}, \label{eq:eToP}
\end{equation}
which immediately follows from Eq.~\eqref{eq:FWgInvRel}.
The radial unit vector is then given by
\begin{equation}
	n^a = \frac{x^a}{r} =  \frac{-e^a_{\alpbar}\sigma^{\alpbar}}{\sqrt{2\sbar}}. \label{eq:nbar}
\end{equation}

Additionally, we must replace the Fermi--Walker basis one-forms, as when written explicitly, the singular field has the standard form
\begin{equation}
	h^{\S}_{\mu\nu}\odif{x^{\mu}}\odif{x^{\nu}} = h^{\S}_{tt}\odif{t}\odif{t} + 2h^{\S}_{ta}\odif{t}\odif{x^a} + h^{\S}_{ab}\odif{x^a}\odif{x^b}. \label{eq:hSFWFull}
\end{equation}
These are given in Eqs.~(82)--(84) from \citetalias{Pound2014} by
\begin{align}
	\odif{t} ={}& \mu \sigma_{\alpbar\alpha}u^{\alpbar}\odif{x^\alpha}, \label{eq:dtFWBar} \\
	\odif{x^a} ={}& -e^a_{\alpbar}(\sigma^{\alpbar}{}_{\alpha} + \mu\sigma^{\alpbar}{}_{\betbar}u^{\betbar}\sigma_{\alpha\bar{\gamma}}u^{\bar{\gamma}}), \label{eq:dxFWBar}
\end{align}
where
\begin{equation}
	\mu = -(\sigma_{\alpbar\betbar}u^{\alpbar}u^{\betbar} + \sigma_{\alpbar}a^{\alpbar})^{-1}. \label{eq:muDef}
\end{equation}
Finally, the second-order singular field \(h^{\S\R}_{\mu\nu}\) features derivatives of the first-order regular field, \(h^{\R1}_{\mu\nu}\).
Using Eqs.~(122)--(123) of \citetalias{Pound2014}, these can be written as
\begin{align}
	\partial_{t}h^{\R1}_{\mu\nu} ={}& h^{\R1}_{\mubar\nubar|\alpbar}u^{\alpbar} + \order{a^\mu}, \label{eq:pdthR1Exp} \\
	\partial_{a}h^{\R1}_{\mu\nu} ={}& h^{\R1}_{\mubar\nubar|\alpbar}e^{\alpbar}_{a} + \order{a^\mu}, \label{eq:pdahR1Exp} \\
	\partial_{t}\partial_{t}h^{\R1}_{\mu\nu} ={}& h^{\R1}_{\mubar\nubar|\alpbar\betbar}u^{\alpbar}u^{\betbar} + \order{a^\mu}, \label{eq:pdtpdthR1Exp} \\
	\partial_{t}\partial_{a}h^{\R1}_{\mu\nu} ={}& h^{\R1}_{\mubar\nubar|\alpbar\betbar}e^{\alpbar}_{a}u^{\betbar} + \order{a^\mu}, \label{eq:pdtpdahR1Exp} \\
	\partial_{a}\partial_{b}h^{\R1}_{\mu\nu} ={}& h^{\R1}_{\mubar\nubar|\alpbar\betbar}e^{\alpbar}_{a}e^{\betbar}_{b} + 2R^{\mubar}{}_{bta}u_{(\alpbar}h^{\R1}_{\betbar)\mubar} \nonumber \\
		& - \frac{4}{3}R^{\mubar}{}_{(b\nubar)a}P^{\nubar}{}_{(\alpbar}h^{\R1}_{\betbar)\mubar} + \order{a^\mu}, \label{eq:pdapdbhR1Exp}
\end{align}
where the bar, \(|\), indicates a covariant derivative at \(x^{\alpbar}\) and any acceleration terms can be ignored as they would belong to the third-order singular field.
These expressions can be derived by taking covariant derivatives of \(h^{\R1}_{\alpbar\betbar}\) and calculating the Christoffel symbols constructed from the FW background metric in Eq.~\eqref{eq:FWmetric}.

After rewriting all quantities in terms of \(\xbar\), we then re-expand them in powers of \(\Delta\tau\), the time difference given in Eq.~\eqref{eq:deltatau}.
For example,
\begin{equation}
	h_{tt}(x,\xbar) = \sum_{n=0}^{\infty}\Delta\tau^n\odv*[order={n}]{h_{tt}(x,x')}{\tau'}, \label{eq:httExp}
\end{equation}
where \(\odv{}{{\tau'}} = u^{\alpha'}\nabla_{\alpha'}\) and the expansion in distance of the difference in proper time is given by
\begin{equation}
	\Delta\tau = \lambda\rb + \lambda^2 \rb a_\sigma + \order{\lambda^3}, \label{eq:deltatauexp}
\end{equation}
originally from Eqs.~(97)--(98) in \citetalias{Pound2014}.
Here, \(\lambda\) is our formal order-counting parameter from Sec.~\ref{sec:CovExp}, and we have introduced the quantity,
\begin{equation}
    \rb \coloneqq u_{\mu'}\sigma^{\mu'}, \label{eq:rdef}
\end{equation}
and below we will also use the quantity,
\begin{equation}
    \rhob \coloneqq \sqrt{P_{\mu'\nu'}\sigma^{\mu'}\sigma^{\nu'}}. \label{eq:rhodef}
\end{equation}
for notational simplicity.\footnote{We use \(\rb\) in agreement with Refs.~\cite{Pound2014,Haas2006,Heffernan2012} but we use \(\rhob\) to match Refs.~\cite{Warburton2014,WardellWarburton2015} instead of \(\mathsf{s}\) as in \citetalias{Pound2014}.}
This means that the contraction of Synge's world function with itself can be written as
\begin{equation}
    \sigma^{\mu'}\sigma_{\mu'} = 2\sigma(x,x') = \rhob^2 - \rb^2. \label{eq:sigmaDSR}
\end{equation}
Here, \(\rb\) gives a notion of the difference in proper time while \(\rhob\) denotes a difference in proper distance.

We note that we expand all quantities (such as Eqs.~\eqref{eq:httExp}--\eqref{eq:deltatauexp}) through four total orders, but we only display the leading two orders here to indicate the forms of the expressions; the full expansions can be found in \citetalias{Pound2014}.
We may do our series expansions as a normal power series as all the Fermi--Walker quantities (including one-forms) are scalars at \(\xbar\).
The expansion of Synge's world function is given by Eqs.~(99)--(101) of \citetalias{Pound2014} as
\begin{align}
	\sigma(x,\xbar) ={}& \sigma(x,x') + \odv{\sigma}{\tau'}\Delta\tau + \frac{1}{2}\odv[order={2}]{\sigma}{{\tau'}}\Delta\tau^2 + \frac{1}{6}\odv[order={3}]{\sigma}{{\tau'}}\Delta\tau^3 \nonumber \\
		& + \order{\lambda^4} \nonumber \\
	={}& \frac{1}{2}\bigl[\lambda^2\rhob^2 + \lambda^3\rb^2 a_\sigma\bigr] + \order{\lambda^4}, \label{eq:sigPrimeExp}
\end{align}
and expansions of the Fermi--Walker basis one-forms are then given by Eqs.~(103)--(106) of \citetalias{Pound2014} as
\begin{align}
	\odif{t} ={}& -g^{\alpha'}_\mu\bigl[\lambda^0 u_{\alpha'} + \lambda (\rb a_{\alpha'} + a_{\sigma}u_{\alpha'}) + \order{\lambda^2}\bigr]\odif{x^\mu}, \label{eq:dt1Form} \\
	\odif{x^a} ={}& g^{\alpha'}_{\mu}\bigl[\lambda^0 e^a_{\alpha'} + \lambda (e^{a\beta'}\rb u_{\alpha'}a_{\beta'}) + \order{\lambda^2}\bigr]\odif{x^\mu}. \label{eq:dx1Form}
\end{align}

In the above expressions, we see that acceleration terms have appeared.
This is a result of taking the derivatives with respect to \(\tau'\).
As stated, \(\odv{}/{{\tau'}} = u^{\alpha'}\nabla_{\alpha'}\), so taking multiple \(\tau'\) derivatives results in us taking derivatives of \(u^{\alpha'}\) along the worldline, providing us with acceleration terms.
These can then be differentiated along the worldline, giving us terms like \(\dot{a}^{\alpha'}\), where a dot indicates a time derivative in the usual manner.

When accounting for these terms, at first order, we split up \(h^{\S1}_{\mu\nu}\) into an acceleration-independent and a linear-in-acceleration piece:
\begin{equation}
	h^{\S1}_{\mu\nu} = h^{\S1\acan}_{\mu\nu} + h^{\S1\abold}_{\mu\nu} + \order{a^2}. \label{eq:hS1AccSplit}
\end{equation}
Recall from Eqs.~\eqref{eq:geodesicForced} and~\eqref{eq:2ndorderselfforce} that each acceleration term carries an \(\epsilon\).
This effectively makes \(h^{\S1\abold}_{\mu\nu}\) a second-order term and allows us to ignore any non-linear acceleration terms that appear in the expansion of \(h^{\S1}_{\mu\nu}\).
Additionally, we can ignore any explicit acceleration terms that appear in both \(h^{\S\R}_{\mu\nu}\) and \(h^{\S\S}_{\mu\nu}\) as these would become third-order terms.

\section{Creating the covariant puncture}\label{sec:HRCov}

With the methods from \citetalias{Pound2014} recapped, we can now proceed to use them to generate our covariant puncture in the highly regular gauge.
We begin in Sec.~\ref{sec:HRMetricPerts} by reviewing the form of the metric perturbations in the highly regular gauge.
Section~\ref{sec:HRCovXBar} will provide the components of the highly regular gauge singular field when evaluated at \(\xbar\) with each being written in covariant form.
We then move to Sec.~\ref{sec:HRCovXPrime}, which provides the components evaluated at \(x'\) before combining this with one-form expansions to find the final, fully covariant form in Sec.~\ref{sec:HRCovPuncFinal}.

\subsection{Metric perturbations in the highly regular gauge}\label{sec:HRMetricPerts}

In this section, we review the main results from \citetalias{Upton2021}.
All results in this section are from there but are reproduced here for convenience.

We write the metric perturbations in the highly regular gauge as
\begin{equation}
    \fullg_{\mu\nu} = \geff_{\mu\nu} + h^{\S}_{\mu\nu}, \label{eq:metricHR}
\end{equation}
where the singular field is given by
\begin{equation}
    h^{\S}_{\mu\nu} = \e h^{\S1}_{\mu\nu} + \e^{2}h^{\S2}_{\mu\nu} + \order{\e^3}. \label{eq:HRsing}
\end{equation}
The second-order singular field is then split as
\begin{equation}
    h^{\S2}_{\mu\nu} = h^{\S\S}_{\mu\nu} + h^{\S\R}_{\mu\nu}, \label{eq:hSSSR}
\end{equation}
where the ``singular times singular'' piece, \(h^{\S\S}_{\mu\nu}\), features all terms proportional to \(m^2\) and the ``singular times regular'' piece, \(h^{\S\R}_{\mu\nu}\), features all terms with the form \(m h^{\R1}_{\mu\nu}\).

The full expressions for the first-order singular field in the highly regular gauge are given in Eq.~(56) of \citetalias{Upton2021}.
We reproduce the two leading orders here:
\begin{subequations}
\label{eq:hS1Prime}
\begin{align}
    h^{\S1}_{tt} ={}& \frac{2m}{r} + \frac{11}{3}mr\E_{ab}\nhat^{ab}  + \order{r^2}, \label{eq:hS1Primett} \displaybreak[0] \\
    h^{\S1}_{ta} ={}& \frac{2m}{r}\nhat_a + \frac{2}{15}mr\big(11\E_{ab}\nhat^{b} + 10\B^{bc}\epsilon_{acd}\nhat_{b}{}^{d} \nonumber \\
        & + 15\E_{bc}\nhat_{a}{}^{bc}\big) + \order{r^2}, \label{eq:hS1Primeta} \displaybreak[0] \\
    h^{\S1}_{ab} ={}&  \frac{2m}{3r}\big(\delta_{ab} + 3\nhat_{ab}\big) + \frac{1}{315}mr\big(154\E_{ab} \nonumber \\
        & - 168\B^{d}_{(a}\epsilon_{b)cd}\nhat^{c} + 580\E^{c}{}_{(a}\nhat_{b)c} + 15\E_{cd}\delta_{ab}\nhat^{cd} \nonumber \\
        & + 840\B^{cd}\epsilon_{c}{}^{i}{}_{(a}\nhat_{b)di} + 105\E_{cd}\nhat_{ab}{}^{cd}\big) + \order{r^2}. \label{eq:hS1Primeab}
\end{align}%
\end{subequations}
Moving to second order, \(h^{\S\R}_{\mu\nu}\) is given in full by Eq.~(130) of \citetalias{Upton2021}.
The two leading orders are
\begin{subequations}
\label{eq:hrgaugeSR}
\begin{align}
    h^{\mathrm{SR}}_{tt} ={}& -\frac{m}{r}\big(2 h^{\R1}_{tt} + h^{\R1}_{ab} n^{ab}\big) - \frac{mr^0}{2} \bigl(h^{\R1}_{ab,c} n^{abc} \nonumber \\
        & - 4n^{ab} \partial_{t}h^{\R1}_{ab} + 8 n^{a} \partial_{t}h^{\R1}_{ta}\bigr) + \order{r}, \label{eq:hrgaugeSRtt} \displaybreak[0] \\
    h^{\mathrm{SR}}_{ta} ={}& -\frac{m}{r}\bigl(2 h^{\R1}_{ta} + h^{\R1}_{tt} n_{a} - 2 h^{\R1}_{ab} n^{b} + 2 h^{\R1}_{bc} n_{a}{}^{bc}\bigr) \nonumber \\
        & - \frac{mr^{0}}{2}\big(2n^b\big[2 h^{\R1}_{t[a,b]} - \partial_{t}h^{\R1}_{ab}\big] + 4n_a{}^{b}\partial_{t}h^{\R1}_{tb} \nonumber \\
        & - n^{bc}\big[2\,^{0}{}h^{\R1}_{ab,c} + h^{\R1}_{bc,a}\big] - 2n_a{}^{bc}\partial_{t}h^{\R1}_{bc}\bigr) + \order{r}, \label{eq:hrgaugeSRta} \displaybreak[0] \\
    h^{\mathrm{SR}}_{ab} ={}& -\frac{m}{r}\bigl(4 h^{\R1}_{t(a} n_{b)} - 4 h^{\R1}_{c(a} n_{b)}{}^{c} + 3 h^{\R1}_{cd} n_{ab}{}^{cd}\bigr) \nonumber \\
        & - \frac{mr^0}{2}\bigl(4 h^{\R1}_{t(a,|c|} n_{b)}{}^{c} + 4 h^{\R1}_{tc,(a} n_{b)}{}^{c} \nonumber \\
        & - 4 h^{\R1}_{c(a,|d|} n_{b)}{}^{cd} -  2 h^{\R1}_{cd,(a} n_{b)}{}^{cd} \nonumber \\
        & + 3 h^{\R1}_{cd,i} n_{ab}{}^{cdi} - 4n^{c}{}_{(a} \partial_{|t|}h^{\R1}_{b)c}\bigr) + \order{r}. \label{eq:hrgaugeSRab}
    \end{align}%
\end{subequations}
Finally, \(h^{\S\S}_{\mu\nu}\) is given by Eq.~(131) of \citetalias{Upton2021}, which we reproduce here in full as
\begin{subequations}
\label{eq:hrgaugeSS}
\begin{align}
    h^{\mathrm{SS}}_{tt} ={}& -4m^2\big[r^0\mathcal{E}_{ab} n^{ab} + r\big(\tfrac{1}{3} \dot{\mathcal{E}}_{ab} n^{ab} \big\{11 - 6 \log(\tfrac{2 m}{r})\big\} \nonumber \\
        & + \tfrac{2}{3} \mathcal{E}_{abc} n^{abc}\big)\big] + \order{r^2}, \label{eq:hrgaugeSStt} \displaybreak[0] \\
    h^{\mathrm{SS}}_{ta} ={}& - 4m^2\big[r^0\mathcal{E}_{bc} n_{a}{}^{bc} + r\big(\tfrac{2}{9} \dot{\mathcal{E}}_{ab} n^{b} \big\{7 - 3 \log(\tfrac{2 m}{r})\big\} \nonumber \\
        & + \tfrac{1}{6} \mathcal{E}_{abc} n^{bc} - \tfrac{2}{9} \dot{\mathcal{B}}_{b}{}^{d} \epsilon_{acd} n^{bc} \big\{4 - 3 \log(\tfrac{2 m}{r})\big\} \nonumber \\
        & + \tfrac{1}{9} \dot{\mathcal{E}}_{bc} n_{a}{}^{bc} \big\{19 - 12 \log(\tfrac{2 m}{r})\big\} + \tfrac{1}{2} \mathcal{E}_{bcd} n_{a}{}^{bcd} \nonumber \\
        & - \tfrac{2}{9} \mathcal{B}_{bc}{}^{i} \epsilon_{adi} n^{bcd}\big)\big] + \order{r^2}, \label{eq:hrgaugeSSta} \displaybreak[0] \\
    h^{\mathrm{SS}}_{ab} ={}& - 4m^2\bigl[r^0\bigl(\mathcal{B}_{(a}{}^{d} \epsilon_{b)cd} n^{c} - \tfrac{1}{3} \mathcal{E}_{ab} + \tfrac{2}{3} \mathcal{E}_{c(a} n_{b)}{}^{c} \nonumber \\
        & -  \tfrac{1}{6} \mathcal{E}_{cd} \delta_{ab} n^{cd} -  \mathcal{B}_{c}{}^{i} \epsilon_{di(a} n_{b)}{}^{cd} + \tfrac{5}{6} \mathcal{E}_{cd} n_{ab}{}^{cd}\bigr) \nonumber \\
        & + r\big(\tfrac{2}{3} \dot{\mathcal{E}}_{ab} + \tfrac{4}{9} \dot{\mathcal{E}}_{c(a} n_{b)}{}^{c} \big\{4 - 3 \log(\tfrac{2 m}{r})\big\} \nonumber \\
        & + \tfrac{1}{3} \dot{\mathcal{E}}_{cd} \delta_{ab} n^{cd} +\tfrac{1}{9}n^{cd}{}_{(a}\big\{3 \mathcal{E}_{b)cd}  - 4 \dot{\mathcal{B}}_{|c|}{}^{i} \epsilon_{b)di} \nonumber \\
        & \times \big[4 - 3 \log(\tfrac{2 m}{r})\big]\big\} - \tfrac{4}{9} \mathcal{B}_{cd}{}^{j} \epsilon_{ij(a} n_{b)}{}^{cdi} \nonumber \\
        & + \tfrac{2}{9} \dot{\mathcal{E}}_{cd} n_{ab}{}^{cd} \big\{4 - 3 \log(\tfrac{2 m}{r})\big\} + \tfrac{1}{3} \mathcal{E}_{cdi} n_{ab}{}^{cdi} \big)\bigr] \nonumber \\
        & + \order{r^2}. \label{eq:hrgaugeSSab}
    \end{align}%
\end{subequations}

\subsection{Perturbation components expanded about \texorpdfstring{\(\xbar\)}{x̅}}\label{sec:HRCovXBar}

We begin by calculating the form of the components of the first-order singular field, \(h^{\S1}_{\mu\nu}\), when expanded around \(\xbar^{\alpha}\).
To do so, we substitute the appropriate expressions from Sec.~\ref{sec:FWtoCov} into Eq.~\eqref{eq:hS1Prime}.
The components of \(h^{\S1}_{\mubar\nubar}\) are then given by
\begin{subequations}
\label{eq:hS1xbar}
\begin{align}
	h^{\S1}_{tt} ={}& \frac{\sqrt{2}m}{\lambda\sqrt{\sbar}} + \frac{11m\lambda}{3\sqrt{2\sbar}}R_{\ubar\sbar\ubar\sbar} + \order{\lambda^2}, \label{eq:hS1ttxbar} \displaybreak[0] \\
	h^{\S1}_{ta} ={}& -\frac{m e^\alpbar_a}{3\sbar}\biggl(\frac{3\sigma_\alpbar}{\lambda} - \lambda\Bigl[2\sqrt{2}R_{\alpbar\sbar\ubar\sbar}\sqrt{\sbar}-2R_{\alpbar\ubar\sbar\ubar}\sbar \nonumber \\
		& - 3R_{\ubar\sbar\ubar\sbar}\sigma_{\alpbar}\Bigr]\biggr) + \order{\lambda^2}, \label{eq:hS1taxbar} \displaybreak[0] \\
	h^{\S1}_{ab} ={}& \frac{m e^\alpbar_{a}e^\betbar_{b}}{12\sbar^{3/2}}\biggl(\frac{6\sqrt{2}\sigma_{\alpbar}\sigma_{\betbar}}{\lambda} + \lambda\Bigl[\sqrt{2}R_{\ubar\sbar\ubar\sbar}\sigma_{\alpbar}\sigma_{\betbar} \nonumber \\
		& - 16\sqrt{\sbar}\sigma_{(\alpbar}R_{\betbar)\sbar\ubar\sbar} + 8\sqrt{2}\sbar\sigma_{(\alpbar}R_{\betbar)\ubar\sbar\ubar}\Bigr]\biggr) + \order{\lambda^2}. \label{eq:hS1abxbar}
	\end{align}
\end{subequations}
We have omitted the highest-order piece of \(h^{\S1}_{\mubar\nubar}\) due to its length, but it will be used to calculate the covariant punctures.

This can then be continued at second order for the singular fields \(h^{\S\R}_{\mubar\nubar}\)~\eqref{eq:hrgaugeSR} and \(h^{\S\S}_{\mubar\nubar}\)~\eqref{eq:hrgaugeSS}.
The ``singular times regular'' piece is given by
\begin{subequations}
\label{eq:hSRxbar}
\begin{align}
	h^{\S\R}_{tt} ={}& -\frac{m}{4\sqrt{2}\sbar^{3/2}} \biggl[\frac{2}{\lambda} \Bigl(h^{\R 1}_{\bar{\sigma} \bar{\sigma}} + 4 h^{\R 1}_{\bar{u} \bar{u}} \bar{\sigma}\Bigr) - \lambda^0\Bigl(16 \bar{\sigma} \dot{h}^{\R 1}_{\sbar\ubar}{} \nonumber \\
		& + h^{\R 1}_{\sbar\sbar|\bar{\sigma}} + 4 \sqrt{2} \bar{\sigma}^{1/2} \dot{h}^{\R 1}_{\sbar\sbar}{}\Bigr)\biggr] + \order{\lambda}, \label{eq:hSRttxbar} \displaybreak[0] \\
	h^{\S\R}_{ta} ={}& - \frac{m e_{a}^{\bar{\alpha}}}{4 \bar{\sigma}^2}\biggl[\frac{2}{\lambda}\Bigl(2 h^{\R 1}_{\bar{\alpha} \bar{\sigma}} \bar{\sigma} + 2 \sqrt{2} h^{\R 1}_{\bar{\alpha} \bar{u}} \bar{\sigma}^{3/2} -  h^{\R 1}_{\bar{\sigma} \bar{\sigma}} \sigma_{\bar{\alpha}} \nonumber \\
		& -  h^{\R 1}_{\bar{u} \bar{u}} \bar{\sigma} {\sigma}_{\bar{\alpha}}\Bigr) + \lambda^0\Bigl(\bar{\sigma}_{\bar{\alpha}} h^{\R 1}_{\sbar\sbar|\bar{\sigma}} - 2 \sqrt{2} \bar{\sigma}^{3/2} \bigl(h^{\R 1}_{\sbar\ubar|\bar{\alpha}} \nonumber \\
		& + h^{\R 1}_{\bar{\alpha} \ubar|\bar{\sigma}} - \dot{h}^{\R 1}_{\bar{\alpha} \sbar}{}\bigr) +  \sqrt{2\sbar} {\sigma}_{\bar{\alpha}} \dot{h}^{\R 1}_{\sbar\sbar} - \bar{\sigma} \bigl(h^{\R 1}_{\sbar\sbar|\bar{\alpha}} \nonumber \\
		& + 2 h^{\R 1}_{\bar{\alpha} \sbar|\bar{\sigma}} - 4 {\sigma}_{\bar{\alpha}} \dot{h}^{\R 1}_{\sbar \ubar}\bigr)\Bigr)\biggr] + \order{\lambda}, \label{eq:hSRtaxbar} \displaybreak[0] \\
	h^{\S\R}_{ab} ={}& -\frac{m e_{a}^{\bar{\alpha}} e_{b}^{\bar{\beta}}}{16 \bar{\sigma}^{5/2}} \biggl[\frac{2}{\lambda}\Bigl(3 \sqrt{2} h^{\R 1}_{\bar{\sigma} \bar{\sigma}} {\sigma}_{\bar{\alpha}} {\sigma}_{\bar{\beta}} - 8 \sqrt{2} \bar{\sigma} h^{\R 1}_{\bar{\sigma} (\bar{\alpha}}{\sigma}_{\bar{\beta})} \nonumber \\
		& - 16 \bar{\sigma}^{3/2} h^{\R 1}_{\bar{u}(\bar{\alpha}}{\sigma}_{\bar{\beta})}\Bigr) + \lambda^0\Bigl(4\sqrt{2}\sbar\bigl(h^{\R1}_{\sbar\sbar|(\alpbar}\sigma_{\betbar)} \nonumber \\
		& + 2\sigma_{(\alpbar}h^{\R1}_{\betbar)\sbar|\sbar}\bigr) - 3\sqrt{2}\sigma_{\alpbar}\sigma_{\betbar}h^{\R1}_{\sbar\sbar|\sbar} + 16\sbar^{3/2} \nonumber \\
		& \times \bigl(h^{\R1}_{\sbar\ubar|(\alpbar}\sigma_{\betbar)} + \sigma_{(\alpbar}h^{\R1}_{\betbar)u|\sbar} - \sigma_{(\alpbar}\dot{h}^{\R1}_{\betbar)\sbar}\bigr)\Bigr)\biggr] + \order{\lambda}. \label{eq:hSRabxbar}
\end{align}%
\end{subequations}
As in the expression for \(h^{\S1}_{\mubar\nubar}\), we omit the highest-order piece of \(h^{\S\R}_{\mubar\nubar}\) due to length constraints.
Finally, the ``singular times singular'' piece is given by
\begin{subequations}
\label{eq:hSSxbar}
\begin{align}
	h^{\S\S}_{tt} ={}& - \frac{2m^2\lambda^0}{\sbar} + \frac{2m^2\lambda}{3\sbar}\biggl[2R_{\ubar\sbar\ubar\sbar|\sbar} - \sqrt{2\sbar}\Rdot_{\ubar\sbar\ubar\sbar} \nonumber \\
		& \times \Bigl\{11-6\log(\tfrac{\sqrt{2}m}{\lambda\sqrt{\sbar}})\Bigr\}\biggr] + \order{\lambda^2}, \label{eq:hSSttxbar} \displaybreak[0] \\
	h^{\S\S}_{ta} ={}& \frac{m^2 e_{a}^{\bar{\alpha}}}{18 \bar{\sigma}^{3/2}} \biggl[18 \sqrt{2}\lambda^0 R_{\ubar\sbar \ubar\sbar}{} \sigma_{\bar{\alpha}} - \lambda \Bigl(2 \sqrt{2} \bar{\sigma}\Bigl[2 R_{\bar{\alpha}\ubar\sbar \ubar| \sigma}{} \nonumber \\
		& +  R_{\ubar\sbar \ubar\sbar |\bar{\alpha}}{} + 4 \dot{R}_{\bar{\alpha}\sigma \ubar\sbar}{}\Bigl\{4 - 3 \log(\tfrac{\sqrt{2} m}{\lambda \sqrt{\bar{\sigma}}})\Bigr\}\Bigr] \nonumber \\
		& - 4 \dot{R}_{\bar{\alpha}\ubar\sbar \ubar}{} \bar{\sigma}^{3/2}\Bigl\{29 - 12 \log(\tfrac{\sqrt{2} m}{\lambda \sqrt{\sbar}})\Bigr\} \nonumber \\
		& + 9 \sqrt{2} R_{\ubar\sbar \ubar\sbar | \sigma}{} \sigma_{\bar{\alpha}} - \sqrt{\sbar} \Bigl[6 R_{\bar{\alpha}\sigma \ubar\sbar | \sigma}{} + 2 \dot{R}_{\ubar\sbar \ubar\sbar}{} \nonumber \\
		& \times \Bigl\{37 - 24 \log(\tfrac{\sqrt{2} m}{\lambda \sqrt{\sbar}})\Bigr\} \sigma_{\bar{\alpha}}\Bigr]\Bigr)\biggr] + \order{\lambda^2}, \label{eq:hSStaxbar} \displaybreak[0] \\
	h^{\S\S}_{ab} ={}& \frac{m^2 e^{\alpbar}_{a}e^{\betbar}_{b}}{18\sbar^2} \biggl[3\lambda^0 \Bigl(2 R_{\ubar\sbar \ubar\sbar}{} g_{\bar{\alpha} \bar{\beta}} \bar{\sigma} + 8 R_{\bar{\alpha}\ubar \bar{\beta} \ubar}{} \bar{\sigma}^2 \nonumber \\
		& - 5 R_{\ubar\sbar \ubar\sbar}{} \sigma_{\bar{\alpha}} \sigma_{\bar{\beta}} - 12 \sqrt{2} \bar{\sigma}^{3/2} R_{\ubar (\bar{\alpha} \bar{\beta}) \sigma}{} + 6 \sqrt{2} \bar{\sigma}^{1/2} \nonumber \\
		& \times R_{\sigma \ubar\sbar(\bar{\alpha}}{}\sigma_{\bar{\beta})} - 8 \bar{\sigma} R_{\ubar\sbar \ubar(\bar{\alpha}}{}\sigma_{\bar{\beta})}\Bigr) + \lambda \Bigl(6 R_{\ubar\sbar \ubar\sbar | \sigma}{} \nonumber \\
		& \times \sigma_{\bar{\alpha}} \sigma_{\bar{\beta}} - 48 \sqrt{2} \dot{R}_{\bar{\alpha}\ubar \bar{\beta}\ubar}{} \bar{\sigma}^{5/2} - \bar{\sigma}^{1/2} \Bigl[2 \sqrt{2} \dot{R}_{\ubar\sbar \ubar\sbar}{} \nonumber \\
		& \times \Bigl\{7 - 6 \log(\tfrac{\sqrt{2}m}{\lambda\sqrt{\sbar}})\Bigr\} \sigma_{\bar{\alpha}} \sigma_{\bar{\beta}} + 6 \sqrt{2} \sigma_{(\bar{\alpha}}R_{\bar{\beta})\sigma \ubar\sbar | \sigma}{}\Bigr] \nonumber \\
		& + \bar{\sigma} \Bigl[\Bigl\{64 - 48 \log(\tfrac{\sqrt{2}m}{\lambda\sqrt{\sbar}})\Bigr\} \dot{R}_{\sigma \ubar\sbar(\bar{\alpha}}\sigma_{\bar{\beta})} \nonumber \\
		& + 4 \Bigl(2 \sigma_{(\bar{\alpha}}R_{\bar{\beta})\ubar\sbar \ubar| \sigma}{} + R_{\ubar\sbar \ubar\sbar |(\bar{\alpha}}{}\sigma_{\bar{\beta})}\Bigr)\Bigr] - 4 \sqrt{2} \bar{\sigma}^{3/2} \nonumber \\
		& \times \Bigl[3 \dot{R}_{\ubar\sbar \ubar\sbar}{} g_{\bar{\alpha} \bar{\beta}} + \Bigl\{17 - 12 \log(\tfrac{\sqrt{2}m}{\lambda\sqrt{\sbar}})\Bigr\} \nonumber \\
		& \times \sigma_{(\bar{\alpha}}\dot{R}_{\bar{\beta})\ubar\sbar \ubar}{}\Bigr]\Bigr)\biggr] + \order{\lambda^2}. \label{eq:hSSabxbar}
	\end{align}
\end{subequations}

\subsection{Expansion about \texorpdfstring{\(x'\)}{x'}}\label{sec:HRCovXPrime}

Accounting for the introduction of acceleration terms and splitting up \(h^{\S1}_{\mu\nu}\) as in Eq.~\eqref{eq:hS1AccSplit}, we find that the components of \(h^{\S1\acan}_{\mu\nu}\), when expanded around \({x'}^\alpha\), are given by
\begin{subequations}
\label{eq:hS1NoAccxprime}
\begin{align}
	h^{\S1\acan}_{tt} ={}& \frac{2 m}{\lambda \rhob} + \frac{m \lambda}{3 \rhob^3} R_{u\sigma u\sigma}{} (\rb^2 + 11 \rhob^2) + \order{\lambda^2}, \label{eq:hS1NoAccttxprime} \displaybreak[0] \\
	h^{\S1\acan}_{ta} ={}& - \frac{m e_{a}^{\alpha '}}{36 \rhob^4}\biggl[\frac{72 \rhob^2 \sigma_{\alpha '}}{\lambda} + \lambda \Bigl(12 \rhob^2 \bigl(R_{\alpha 'u\sigma u}{} \nonumber \\
		& \times (\rb^2 + 4 \rb \rhob + 2 \rhob^2) - 2 R_{\alpha '\sigma u\sigma}{} (\rb + 2 \rhob)\bigr) \nonumber \\
		& + 24 R_{u\sigma u\sigma}{} (\rb^2 + 3 \rhob^2) \sigma_{\alpha '}\Bigr)\biggr] + \order{\lambda^2}, \label{eq:hS1NoAcctaxprime} \displaybreak[0] \\
	h^{\S1\acan}_{ab} ={}& \frac{m e_{a}^{\alpha'}e_{b}^{\beta'}}{3\rhob^5}\biggl[\frac{6 \rhob^2 \sigma_{\alpha '} \sigma_{\beta '}}{\lambda} + \lambda\Bigl(R_{u\sigma u\sigma}{} \sigma_{\alpha '} \sigma_{\beta '} \nonumber \\
		& \times (3 \rb^2 + \rhob^2) - 2 \rhob^2 \Bigl[2 (\rb + 2 \rhob) \sigma_{(\alpha'}R_{\beta')\sigma u\sigma}{} \nonumber \\
		& -  (\rb^2 + 4 \rb \rhob + 2 \rhob^2) \sigma_{(\alpha '}R_{\beta ')u\sigma u}{}\Bigr]\Bigr)\biggr] + \order{\lambda^2}. \label{eq:hS1NoAccabxprime}
\end{align}
\end{subequations}
The acceleration terms that appear as a result of our expansion of the first-order singular field are
\begin{subequations}
\label{eq:hS1Accxprime}
\begin{align}
	h^{\S1\abold}_{tt} ={}& - \frac{m\lambda^{0} a_{\sigma}{} \rb^2}{\rhob^3} -  \frac{m \lambda \dot{a}_{\sigma}{} \rb^3}{3 \rhob^3} + \order{\lambda^2}, \label{eq:hS1Accttxprime} \\
	h^{\S1\abold}_{ta} ={}& -\frac{m e_{a}^{\alpha '} \rb}{3 \rhob^6}\bigl[3\lambda^0 \rb \rhob^2 (a_{\alpha '} \rhob^2 - 2 a_{\sigma}{} \sigma_{\alpha '}) \nonumber \\
		& + \lambda \rb^2 \rhob^2 (\dot{a}_{\alpha '} \rhob^2 - 2 \dot{a}_{\sigma}{} \sigma_{\alpha '})\bigr] + \order{\lambda^2}, \label{eq:hS1Acctaxprime} \\
	h^{\S1\abold}_{ab} ={}& - \frac{m e_{a}^{\alpha '} e_{b}^{\beta '} \rb^2}{3 \rhob^5}\bigl[3\lambda^0 (3 a_{\sigma}{} \sigma_{\alpha '} \sigma_{\beta '} - 2 \rhob^2 a_{(\alpha '}\sigma_{\beta ')}) \nonumber \\
		& + \lambda \rb (3 \dot{a}_{\sigma}{} \sigma_{\alpha '} \sigma_{\beta '} - 2 \rhob^2 \dot{a}_{(\alpha '}\sigma_{\beta ')})\bigr] + \order{\lambda^2}. \label{eq:hS1Accabxprime}
	\end{align}
\end{subequations}
As \(h^{\S1\abold}_{\mu\nu}\) is a second-order term, we can neglect any terms of order-\(\lambda^2\) and higher to match the orders required for \(h^{\S\R}_{\mu\nu}\) and \(h^{\S\S}_{\mu\nu}\).

Moving to the second-order field, we calculate the SR components to be
\begin{subequations}
\label{eq:hSRxprime}
\begin{align}
	h^{\S\R}_{tt} ={}& -\frac{m}{2\rhob^3} \Big[\frac{2}{\lambda}\bigl(h^{\R 1}_{\sigma \sigma } + 2 h^{\R 1}_{\sigma u} \rb + h^{\R 1}_{uu} (\rb^2 + 2 \rhob^2)\bigr) \nonumber \displaybreak[0] \\
		& - \lambda^0 \bigl(\rb (\rb h^{\R 1}_{uu;\sigma } + 2 h^{\R 1}_{\sigma u;\sigma }) +  h^{\R 1}_{\sigma \sigma ;\sigma } - (\rb - 4 \rhob) \nonumber \\
		& \times \hdot^{\R 1}_{\sigma \sigma} - 2 (\rb^2 - 4 \rb \rhob - 4 \rhob^2) \hdot^{\R 1}_{\sigma u} \nonumber \\
		& - \rb (\rb^2 - 4 \rb \rhob - 4 \rhob^2) \hdot^{\R 1}_{uu}\bigr)\Big] + \order{\lambda}, \label{eq:hSRttxprime} \displaybreak[0] \\
	h^{\S\R}_{ta} ={}& -\frac{m e_{a}{}^{\alpha '}}{2 \rhob^4} \biggl[\frac{2}{\lambda}\bigl(2 h^{\R 1}_{\alpha ' \sigma } \rhob^2 + 2 h^{\R 1}_{\alpha ' u} \rhob^2 (\rb + \rhob) \nonumber \\
		& -  \bigl(2 h^{\R 1}_{\sigma \sigma } + 4 h^{\R 1}_{\sigma u} \rb + h^{\R 1}_{uu} (2 \rb^2 + \rhob^2)\bigr) \sigma_{\alpha '}\bigr) \nonumber \displaybreak[0] \\
		& - \lambda^0\Bigl[\rhob^2 \Bigl(h^{\R 1}_{\sigma \sigma ;\alpha '} + 2 (\rb + \rhob) h^{\R 1}_{\sigma u;\alpha '} + \rb (\rb + 2 \rhob) \nonumber \\
		& \times h^{\R 1}_{uu;\alpha '} + 2 \bigl(h^{\R 1}_{\alpha ' \sigma ;\sigma } + (\rb + \rhob) (h^{\R 1}_{\alpha ' u;\sigma } -  \hdot^{\R 1}_{\alpha ' \sigma}) \nonumber \\
		& - \rb (\rb + 2 \rhob) \hdot^{\R 1}_{\alpha ' u}\bigr)\Bigr) + \sigma_{\alpha '} \bigl(2 (\rb -  \rhob) \hdot^{\R 1}_{\sigma \sigma} \nonumber \\
		& - 2 (\rb^2 h^{\R 1}_{uu;\sigma } + 2 \rb h^{\R 1}_{\sigma u;\sigma } + h^{\R 1}_{\sigma \sigma ;\sigma }) \nonumber \\
		& + 2 (\rb^2 -  \rb \rhob -  \rhob^2)(2\hdot^{\R 1}_{\sigma u} + \rb \hdot^{\R 1}_{uu})\bigr)\Bigr] \biggr] \nonumber \\
		& + \order{\lambda}, \label{eq:hSRtaxprime} \displaybreak[0] \\
	h^{\S\R}_{ab} ={}& - \frac{m e_{a}^{\alpha '} e_{b}^{\beta '}}{2 \rhob^5}\biggl[\frac{2}{\lambda}\bigl[3 \bigl(h^{\R 1}_{\sigma \sigma } + \rb (2 h^{\R 1}_{\sigma u} + h^{\R 1}_{uu} \rb)\bigr) \sigma_{\alpha '} \sigma_{\beta '} \nonumber \\
		& - 4 \rhob^2 h^{\R 1}_{(\alpha ' |\sigma |}\sigma_{\beta ')} - 4 \rhob^2 (\rb + \rhob) h^{\R 1}_{(\alpha ' |u|}\sigma_{\beta ')}\bigr] \nonumber \displaybreak[0] \\
		& - \lambda^0\Bigl[3 \sigma_{\alpha '} \sigma_{\beta '} \Bigl(h^{\R 1}_{\sigma \sigma ;\sigma } -  \rb \bigl(\hdot^{\R 1}_{\sigma \sigma} -2 h^{\R 1}_{\sigma u;\sigma } \nonumber \\
		& + \rb (2 \hdot^{\R 1}_{\sigma u} - h^{\R 1}_{uu;\sigma } + \rb \hdot^{\R 1}_{uu})\bigr)\Bigr) \nonumber \\
		& - 2 \rhob^2 \Bigl(\sigma_{(\alpha '}h^{\R 1}_{|\sigma \sigma |;\beta ')} + 2 (\rb + \rhob) \sigma_{(\alpha '}h^{\R 1}_{|\sigma u|;\beta ')} \nonumber \\
		& + \rb (\rb + 2 \rhob) \sigma_{(\alpha '}h^{\R 1}_{|uu|;\beta ')} + 2 \bigl(\sigma_{(\alpha '}h^{\R 1}_{\beta ')\sigma;\sigma} \nonumber \\
		& + (\rb + \rhob) (\sigma_{(\alpha '}h^{\R 1}_{\beta ')u;\sigma} -  \sigma_{(\alpha '}\hdot^{\R 1}_{\beta ')\sigma}) \nonumber \\
		& - \rb (\rb + 2 \rhob) \sigma_{(\alpha '}\hdot^{\R 1}_{\beta ')u}\bigr)\Bigr)\Bigr]\biggr] + \order{\lambda}, \label{eq:hSRabxprime}
	\end{align}
\end{subequations}
where, again, we have omitted the highest order term.
The SS components are calculated to be
\begin{subequations}
\label{eq:hSSxprime}
\begin{align}
	h^{\S\S}_{tt} ={}& -\frac{4 m^2}{3 \rhob^2} \Bigl[3\lambda^0 R_{u\sigma u\sigma}{} - \lambda \Bigl(2 R_{u\sigma u\sigma ; \sigma}{} - \Rdot_{u\sigma u\sigma}{} \nonumber \\
		& \times \bigl[\rb + 11 \rhob - 6 \logsr \rhob\bigr]\Bigr)\Bigr] + \order{\lambda^2}, \label{eq:hSSttxprime} \displaybreak[0] \\
	h^{\S\S}_{ta} ={}& \frac{2 m^2 e_{a}^{\alpha '}}{9 \rhob^3} \Bigl[18\lambda^0 R_{u\sigma u\sigma}{} \sigma_{\alpha '} + \lambda \Bigl(\rhob \bigl[3 R_{\alpha '\sigma u\sigma ; \sigma}{} \nonumber \\
		& -  R_{u\sigma u\sigma ;\alpha '}{} \rhob -  R_{\alpha 'u\sigma u; \sigma}{} (3 \rb + 2 \rhob) + \dot{R}_{\alpha '\sigma u\sigma}{} \nonumber \\
		& \times \bigl(3 \rb - 4 \bigl(4 - 3 \logsr\bigr) \rhob\bigr) - \dot{R}_{\alpha 'u\sigma u}{} \nonumber \\
		& \times \bigl(3 \rb^2 - 14 \rb \rhob - 29 \rhob^2 + 12 \logsr \rhob (\rb + \rhob)\bigr)\bigr] \nonumber \\
		& - \sigma_{\alpha '}\bigl[9 R_{u\sigma u\sigma ; \sigma}{} - \dot{R}_{u\sigma u\sigma}{} \bigl(9 \rb + 37 \rhob \nonumber \\
		& - 24 \logsr \rhob\bigr)\bigr]\Bigr)\Bigr] + \order{\lambda^2}, \label{eq:hSStaxprime} \displaybreak[0] \\
	h^{\S\S}_{ab} ={}& \frac{2m^2e^{\alpha'}_{a}e^{\beta'}_{b}}{9\rhob^4}\Bigl[3\lambda^0\Bigl(\rhob^2\bigl[R_{u\sigma u\sigma}{} g_{\alpha ' \beta '} + 2 R_{\alpha 'u \beta ' u}{} \nonumber \\
		& \times \rhob (3 \rb + \rhob) - 6 \rhob R_{u(\alpha ' \beta ') \sigma}{}\bigr] + 2 \rhob  \sigma_{(\alpha '}\bigl[3 R_{\beta ')\sigma u\sigma}{} \nonumber \\
		& -  R_{\beta ')u\sigma u}{} (3 \rb + 2 \rhob)\bigr] - 5 R_{u\sigma u\sigma}{} \sigma_{\alpha '} \sigma_{\beta '}\Bigr) \nonumber \displaybreak[0] \\
		& + \lambda\Bigl(3 \rhob^2 \bigl[\dot{R}_{u\sigma u\sigma}{} g_{\alpha ' \beta '} (\rb - 2 \rhob) + 2 \dot{R}_{\alpha 'u \beta ' u}{} \nonumber \\
		& \times (3 \rb - 2 \rhob) \rhob (\rb + \rhob) - 6 \rb \rhob \dot{R}_{u(\alpha ' \beta ') \sigma}{}\bigr] \nonumber \\
		& + \sigma_{\alpha '} \sigma_{\beta '}\bigl[6 R_{u\sigma u\sigma ; \sigma}{} -  \dot{R}_{u\sigma u\sigma}{} \bigl(9 \rb + 14 \rhob \nonumber \\
		& - 12 \rhob \logsr\bigr)\bigr] - 2 \rhob \bigl[3 \sigma_{(\alpha '}R_{\beta ')\sigma u\sigma ; \sigma}{} \nonumber \\
		& - 2 \bigl(3 \rb + 8 \rhob - 6 \logsr \rhob\bigr) \dot{R}_{\sigma u\sigma(\alpha '}{}\sigma_{\beta ')} \nonumber \\
		& - (3 \rb + 2 \rhob) \sigma_{(\alpha '}R_{\beta ')u\sigma u; \sigma}{} + \bigl(6 \rb^2 + 20 \rb \rhob \nonumber \\
		& + 17 \rhob^2 - 12 \logsr \rhob (\rb + \rhob)\bigr) \dot{R}_{u\sigma u(\alpha '}{}\sigma_{\beta ')} \nonumber \\
		& - \rhob R_{u\sigma u\sigma ;(\alpha '}{}\sigma_{\beta ')}\bigr]\Bigr)\Bigr] + \order{\lambda^2}. \label{eq:hSSabxprime}
	\end{align}
\end{subequations}

\begin{widetext}
\subsection{Final expressions for the covariant punctures}\label{sec:HRCovPuncFinal}

With all of the individual components of the singular field now expressed as functions of \({x'}^\alpha\), we now combine them with the expansions of \(\odif{t}\) and \(\odif{x^{a}}\), given in Eqs.~\eqref{eq:dt1Form}--\eqref{eq:dx1Form} to find the final form of the covariant punctures.
After contracting with the basis vectors, we obtain the covariant form of \(h^{\S}_{\mu\nu}\odif{x^\mu}\odif{x^\nu}\), as in Eq.~\eqref{eq:hSFWFull}.
We then read off the coefficients of \(\odif{x^\mu}\odif{x^\nu}\) to obtain \(h^{\S}_{\mu\nu}\).

The first-order singular field is given by
\begingroup
\allowdisplaybreaks
\begin{align}
	h^{\S1\acan}_{\alpha\beta} ={}& - \frac{m g^{\alpha '}{}_{\alpha } g^{\beta '}{}_{\beta }}{36 \rhob^5}\Biggl[\frac{-72 \rhob^2}{\lambda} \bigl(\sigma_{\alpha '} + (\rb + \rhob) u_{\alpha '}\bigr) \bigl(\sigma_{\beta '} + (\rb + \rhob) u_{\beta '}\bigr) - 12\lambda\Bigl(R_{u\sigma u\sigma}{} (3 \rb^2 + \rhob^2) \sigma_{\alpha '} \sigma_{\beta '} \nonumber \\
		& + 2R_{u\sigma u\sigma}{} \rb (3 \rb -  \rhob) (\rb + \rhob) \sigma_{(\alpha '} u_{\beta ')} + R_{u\sigma u\sigma}{} (\rb -  \rhob) (\rb + \rhob)^2 (3 \rb + \rhob) u_{\alpha '} u_{\beta '} + 2 \rhob^2 \bigl(\sigma_{(\alpha'} + (\rb + \rhob) u_{(\alpha'}\bigr) \nonumber \\
		& \cdot \bigl(R_{\beta')\sigma u\sigma}{} (\rb - 3 \rhob) + 2 R_{\beta') u\sigma u}{} (\rhob^2- \rb^2)\bigr)\Bigr) + \lambda^2\Bigl(3\bigl(-3 \Rdot_{u\sigma u\sigma}{} \rb (\rb^2 + \rhob^2) +  R_{u\sigma u\sigma ; \sigma}{} (3 \rb^2 + \rhob^2)\bigr) \sigma_{\alpha '} \sigma_{\beta '} \nonumber \\
		& + 6\bigl( R_{u\sigma u\sigma ; \sigma}{} \rb (3 \rb -  \rhob) (\rb + \rhob) - \Rdot_{u\sigma u\sigma}{} (3 \rb^4 + 2 \rb^3 \rhob + 3 \rhob^4)\bigr) \sigma_{(\alpha '} u_{\beta ')} + 3\bigl( R_{u\sigma u\sigma ; \sigma}{} (\rb -  \rhob) (\rb + \rhob)^2 (3 \rb + \rhob) \nonumber \\
		& - \Rdot_{u\sigma u\sigma}{} (3 \rb^5 + 4 \rb^4 \rhob - 2 \rb^3 \rhob^2 - 6 \rb^2 \rhob^3 + 3 \rb \rhob^4 + 14 \rhob^5)\bigr) u_{\alpha '} u_{\beta '} + 4 g_{\alpha ' \beta '} \rhob^4 (3 R_{u\sigma u\sigma ; \sigma}{} + 3 \Rdot_{u\sigma u\sigma}{} \rb - 8 \Rdot_{u\sigma u\sigma}{} \rhob) \nonumber \\
		& - 8 \rhob^5 \bigl(\Rdot_{\alpha '\sigma \beta ' \sigma}{} -  R_{\alpha 'u \beta ' u; \sigma}{} (3 \rb + \rhob) + \Rdot_{\alpha 'u \beta ' u}{} (9 \rhob^2 + 4 \rb \rhob - 2 \rb^2) + 3 R_{u (\alpha ' \beta ') \sigma ; \sigma}{} +  (\rb - 5 \rhob) \Rdot_{u (\alpha ' \beta ') \sigma}{} \nonumber \\
		& + 2 \rhob R_{\sigma u u(\alpha';\beta')}{}\bigr) -2 \rhob^2 \sigma_{(\beta '}\bigl(6 (2 \rhob-\rb) R_{\alpha ')\sigma u\sigma ; \sigma}{} + (3 \rb^2 - 24 \rb \rhob - 32 \rhob^2) \Rdot_{\alpha')\sigma u\sigma}{} + 9 \rb^2 R_{\alpha')u\sigma u; \sigma}{} \nonumber \\
		& + 2 \rhob^2 R_{\alpha')u\sigma u; \sigma}{} - 6 \rb^3 \Rdot_{\alpha')u\sigma u}{} + 12 \rb^2 \rhob \Rdot_{\alpha')u\sigma u}{} + 58 \rb \rhob^2 \Rdot_{\alpha')u\sigma u} + 8 \rhob^3 \Rdot_{\alpha')u\sigma u} + \rhob^2 R_{|u\sigma u\sigma|;\alpha')}\bigr) \nonumber \\
		& - 2 \rhob^2 u_{(\beta '}\bigl(6 (4 \rhob^2 + \rb \rhob - \rb^2) R_{\alpha')\sigma u\sigma ; \sigma} + (3 \rb^3 - 21 \rb^2 \rhob - 44 \rb \rhob^2 - 44 \rhob^3) \Rdot_{\alpha')\sigma u\sigma} + 9 \rb^3 R_{\alpha')u\sigma u; \sigma} \nonumber \\
		& + 9 \rb^2 \rhob R_{\alpha')u\sigma u; \sigma} - 10 \rb \rhob^2 R_{\alpha')u\sigma u; \sigma} - 2 \rhob^3 R_{\alpha')u\sigma u; \sigma} - 6 \rb^4 \Rdot_{\alpha')u\sigma u} + 6 \rb^3 \rhob \Rdot_{\alpha')u\sigma u} + 58 \rb^2 \rhob^2 \Rdot_{\alpha')u\sigma u} \nonumber \\
		& + 74 \rb \rhob^3 \Rdot_{\alpha')u\sigma u} + 52 \rhob^4 \Rdot_{\alpha')u\sigma u} + (\rb - 7 \rhob) \rhob^2 R_{|u\sigma u\sigma|;\alpha')}\bigr) + 24 \Rdot_{u\sigma u\sigma}{} \logsr g_{\alpha ' \beta '} \rhob^5 \nonumber \\
		& + 48 \logsr \rhob^4 \bigl(\Rdot_{\alpha ' u \beta ' u} \rhob^2 (\rb + \rhob) - \rhob^2 \Rdot_{u(\alpha ' \beta ') \sigma} - (\sigma_{(\alpha'} + (2\rhob + \rb)  u_{(\alpha'})\Rdot_{\beta')\sigma u\sigma} + \rb \sigma_{(\alpha'}\Rdot_{\beta')u\sigma u} \nonumber \\
		& + (\rb + \rhob)^2 u_{(\alpha'}\Rdot_{\beta ')u\sigma u}\bigr)\Bigr)\Biggr] + \order{\lambda^3}. \label{eq:hS1NoAccCov}
\end{align}%
\endgroup
We have confirmed that this satisfies the Einstein field equations to the appropriate order, i.e.\
\begin{equation}
	\deltaG^{\mu\nu}[h^{\S1\acan}] = \order{\lambda}, \quad x\notin \gamma. \label{eq:hS1NoAccCovEFE}
\end{equation}

At second order, the SS piece of the singular field is given by
\begingroup
\allowdisplaybreaks
\begin{align}
	h^{\S\S}_{\alpha\beta} ={}& - \frac{2 m^2 g^{\alpha '}{}_{\alpha } g^{\beta '}{}_{\beta }}{9 \rhob^4} \Bigl[3\lambda^0\Bigl(R_{u\sigma u\sigma}{} (5 \rb^2 + 6 \rb \rhob + 5 \rhob^2) u_{\alpha '} u_{\beta '} + 5 R_{u\sigma u\sigma}{} \sigma_{\alpha '} \sigma_{\beta '} + 2 R_{u\sigma u\sigma}{} (5 \rb + 3 \rhob) \sigma_{(\alpha '} u_{\beta ')} \nonumber \\
		& - 2\rhob u_{(\alpha '}\bigl(3 R_{\beta ')\sigma u\sigma} \rb - R_{\beta ')u\sigma u}{} (3 \rb^2 + 2 \rb \rhob + 3 \rhob^2)\bigr) - 2 \rhob \sigma_{(\alpha '}\bigl(3 R_{\beta ')\sigma u\sigma}{} -  R_{\beta ')u\sigma u}{} (3 \rb + 2 \rhob)\bigr) \nonumber \\
		& - \rhob^2 \bigl(R_{u\sigma u\sigma}{} g_{\alpha ' \beta '} + 2 R_{\alpha ' u \beta ' u}{} \rhob (3 \rb + \rhob) - 6 \rhob R_{u (\alpha ' \beta ') \sigma}{}\bigr)\Bigr) + \lambda\Bigl(u_{\alpha '} u_{\beta '} \bigl[\dot{R}_{u\sigma u\sigma}{} (\rb + \rhob) \bigl(9 \rb^2 + 11 \rb \rhob + 38 \rhob^2 \nonumber \\
		& - 12 \logsr \rhob (\rb + \rhob) - 6 R_{u\sigma u\sigma ; \sigma}{} (\rb + \rhob)^2\bigr)\bigr] + \sigma_{\alpha '} \sigma_{\beta '}\bigl[\dot{R}_{u\sigma u\sigma}{} \bigl(9 \rb + 14 \rhob - 12 \logsr \rhob\bigr) - 6 R_{u\sigma u\sigma ; \sigma}{}\bigr] \nonumber \\
		& +  2u_{(\alpha '}\sigma_{\beta ')}\bigl[\dot{R}_{u\sigma u\sigma}{} \bigl(9 \rb^2 + 17 \rb \rhob + 20 \rhob^2 - 12 \logsr \rhob (\rb + \rhob)\bigr) - 6 R_{u\sigma u\sigma ; \sigma}{} (\rb + \rhob)\bigr] \nonumber \\
		& - 2\rhob u_{(\alpha '}\Bigl[\dot{R}_{\beta ')\sigma u\sigma}{} \bigl(6 \rb^2 + 13 \rb \rhob + 16 \rhob^2 - 12 \logsr \rhob (\rb + \rhob)\bigr) - 3 R_{\beta ')\sigma u\sigma ; \sigma}{} (\rb + \rhob) + (\rb + \rhob) \nonumber \\
		& \times \bigl(R_{|u\sigma u\sigma |;\beta ')}{} \rhob + R_{\beta ')u\sigma u; \sigma}{} (3 \rb + 2 \rhob) - \dot{R}_{\beta ')u\sigma u}{} \bigl[6 \rb^2 + 11 \rb \rhob + 29 \rhob^2 - 12 \logsr \rhob (\rb + \rhob)\bigr]\bigr)\Bigr] \nonumber \\
		& - 2\rhob \sigma_{(\alpha '}\bigl[R_{|u\sigma u\sigma |;\beta ')}{} \rhob - 3 R_{\beta ')\sigma u\sigma ; \sigma}{} + R_{\beta ')u\sigma u; \sigma}{} (3 \rb + 2 \rhob) + 2 \dot{R}_{\beta ')\sigma u\sigma}{} \bigl(3 \rb + 8 \rhob - 6 \logsr \rhob\bigr) - \dot{R}_{\beta ')u\sigma u}{} \nonumber \\
		& \times \bigl(6 \rb^2 + 20 \rb \rhob + 17 \rhob^2 - 12 \logsr \rhob (\rb + \rhob)\bigr)\bigr] - 3 \rhob^2 \bigl[\dot{R}_{u\sigma u\sigma}{} g_{\alpha ' \beta '} (\rb - 2 \rhob) + 2 \dot{R}_{\alpha 'u \beta ' u}{} (3 \rb - 2 \rhob) \rhob (\rb + \rhob) \nonumber \\
		& - 6 \rb \rhob \dot{R}_{u(\alpha ' \beta ') \sigma}{}\bigr]\Bigr)\Bigr] + \order{\lambda^2}. \label{eq:hSSCov}
\end{align}%
\endgroup
This again satisfies the appropriate Einstein field equations,
\begin{equation}
	\deltaG^{\mu\nu}[h^{\S\S}] + \delta^{2}G^{\mu\nu}[h^{\S1\acan},h^{\S1\acan}] = \order{\lambda^0}, \quad x \notin \gamma. \label{eq:hSSCovEFE}
\end{equation}

The first-order singular field with linear acceleration terms is
\begin{align}
	h^{\S1\abold}_{\alpha\beta} ={}& \frac{g^{\alpha'}{}_{\alpha}g^{\beta'}{}_{\beta}}{3\rhob^5}\Bigl[3\lambda^{0}\Bigl(2 \rb \rhob^2 (\rb + 2 \rhob) a_{(\alpha '} \bigl(\sigma_{\beta ')} + (\rb + \rhob) u_{\beta ')}\bigr) - a_{\sigma}{} \bigl(3 \rb^2 \sigma_{\alpha '} \sigma_{\beta '} + 2(3 \rb^3 + 2 \rb^2 \rhob - 2 \rb \rhob^2 - 2 \rhob^3) \sigma_{(\alpha '} u_{\beta ')} \nonumber \\
        & +  (\rb + \rhob) (3 \rb^3 + \rb^2 \rhob - 4 \rb \rhob^2 - 4 \rhob^3) u_{\alpha '} u_{\beta '}\bigr)\Bigr) + \lambda\Bigl(2 \dot{a}_{(\alpha '} \rb^2 \rhob^2 (\rb + 3 \rhob) \bigl(\sigma_{\beta ')} + (\rb + \rhob) u_{\beta )'}\bigr) - \dot{a}_{\sigma}{} \rb\bigl(3 \rb^2 \sigma_{\alpha '} \sigma_{\beta '} \nonumber \\
        & + 2(3 \rb^3 + 2 \rb^2 \rhob - 3 \rb \rhob^2 - 6 \rhob^3) \sigma_{(\alpha '} u_{\beta ')} +  (\rb + \rhob) (3 \rb^3 + \rb^2 \rhob - 6 \rb \rhob^2 - 12 \rhob^3) u_{\alpha '} u_{\beta '}\bigr)\Bigr)\Bigr] + \order{\lambda^2}, \label{eq:hS1AccCov}
\end{align}
while the SR piece of the second-order singular field is
\begingroup
\allowdisplaybreaks
\begin{align}
	h^{\S\R}_{\alpha\beta} ={}& - \frac{m g^{\alpha '}{}_{\alpha } g^{\beta '}{}_{\beta }}{2 \rhob^5}\biggl[\frac{2}{\lambda}\Bigl[4 \rhob^2 \Bigl(h^{\R 1}_{(\alpha ' |\sigma |}\sigma_{\beta ')} + (\rb + \rhob) \bigl(h^{\R 1}_{(\alpha ' |u|}\sigma_{\beta ')} + h^{\R 1}_{(\alpha ' |\sigma |}u_{\beta ')} + (\rb + \rhob) h^{\R 1}_{(\alpha ' |u|}u_{\beta ')}\bigr)\Bigr) \nonumber \\
		& - h^{\R 1}_{\sigma \sigma } \bigl(3 \sigma_{\alpha '} \sigma_{\beta '} + (\rb + \rhob) (3 \rb + \rhob) u_{\alpha '} u_{\beta '} + 2 (3 \rb + 2 \rhob) \sigma_{(\alpha '}u_{\beta ')}\bigr) - h^{\R 1}_{\sigma u} \Bigl(6 \rb \sigma_{\alpha '} \sigma_{\beta '} + 2 (\rb + \rhob) \bigl((3 \rb - 2 \rhob) \nonumber \\
		& \times (\rb + \rhob) u_{\alpha '} u_{\beta '} + 2 (3 \rb -  \rhob) \sigma_{(\alpha '}u_{\beta ')}\bigr)\Bigr) - h^{\R 1}_{uu} \Bigl(3 \rb^2 \sigma_{\alpha '} \sigma_{\beta '} + (\rb + \rhob) \bigl((\rb + \rhob) (3 \rb^2 - 2 \rb \rhob - 2 \rhob^2) u_{\alpha '} u_{\beta '} \nonumber \\
		& + 2 (3 \rb^2 -  \rb \rhob -  \rhob^2) \sigma_{(\alpha '}u_{\beta ')}\bigr)\Bigr)\Bigr] + \lambda^0\Bigl[h^{\R 1}_{\sigma \sigma ;\sigma } \bigl(-3 \sigma_{\alpha '} \sigma_{\beta '} -  (\rb + \rhob) (3 \rb + \rhob) u_{\alpha '} u_{\beta '} - 2 (3 \rb + 2 \rhob) \sigma_{(\alpha '}u_{\beta ')}\bigr) \nonumber \\
		& + \rb \hdot^{\R 1}_{uu} \bigl(3 \rb^2 \sigma_{\alpha '} \sigma_{\beta '} + (\rb + \rhob) (3 \rb^3 + \rb^2 \rhob - 6 \rb \rhob^2 - 8 \rhob^3) u_{\alpha '} u_{\beta '} + 2 (3 \rb^3 + 2 \rb^2 \rhob - 3 \rb \rhob^2 - 4 \rhob^3) \sigma_{(\alpha '}u_{\beta ')}\bigr) \nonumber \\
		& + \hdot^{\R 1}_{\sigma u} \bigl(6 \rb^2 \sigma_{\alpha '} \sigma_{\beta '} + 2 (\rb + \rhob) (3 \rb^3 + \rb^2 \rhob - 4 \rb \rhob^2 - 4 \rhob^3) u_{\alpha '} u_{\beta '} + 4 (3 \rb^3 + 2 \rb^2 \rhob - 2 \rb \rhob^2 - 2 \rhob^3) \sigma_{(\alpha '}u_{\beta ')}\bigr) \nonumber \\
		& - h^{\R 1}_{uu;\sigma } \bigl(3 \rb^2 \sigma_{\alpha '} \sigma_{\beta '} +  (\rb + \rhob) (3 \rb^3 + \rb^2 \rhob - 4 \rb \rhob^2 - 4 \rhob^3) u_{\alpha '} u_{\beta '} + 2 (3 \rb^3 + 2 \rb^2 \rhob - 2 \rb \rhob^2 - 2 \rhob^3) \sigma_{(\alpha '}u_{\beta ')}\bigr) \nonumber \\
		& - 2 h^{\R 1}_{\sigma u;\sigma } \Bigl(3 \rb \sigma_{\alpha '} \sigma_{\beta '} + (\rb + \rhob) \bigl((3 \rb - 2 \rhob) (\rb + \rhob) u_{\alpha '} u_{\beta '} + 2 (3 \rb -  \rhob) \sigma_{(\alpha '}u_{\beta ')}\bigr)\Bigr) \nonumber \\
		& + \hdot^{\R 1}_{\sigma \sigma} \Bigl(3 \rb \sigma_{\alpha '} \sigma_{\beta '} + (\rb + \rhob) \bigl((3 \rb - 2 \rhob) (\rb + \rhob) u_{\alpha '} u_{\beta '} + 2 (3 \rb -  \rhob) \sigma_{(\alpha '}u_{\beta ')}\bigr)\Bigr) \nonumber \\
		& + 2 \rhob^2 \bigl(\sigma_{(\alpha '}h^{\R 1}_{|\sigma \sigma |;\beta ')} + 2 (\rb + \rhob) \sigma_{(\alpha '}h^{\R 1}_{|\sigma u|;\beta ')} + \rb (\rb + 2 \rhob) \sigma_{(\alpha '}h^{\R 1}_{|uu|;\beta ')} + 2 \sigma_{(\alpha '}h^{\R 1}_{\beta ')\sigma;\sigma } + 2 (\rb+\rhob) \sigma_{(\alpha '}h^{\R 1}_{\beta ')u;\sigma} \nonumber \\
		& - 2 (\rb+\rhob) \sigma_{(\alpha '}\hdot^{\R 1}_{\beta ')\sigma} - 2 \rb(\rb+2\rhob) \sigma_{(\alpha '}\hdot^{\R 1}_{\beta ')u} + (\rb+\rhob) u_{(\alpha '}h^{\R 1}_{|\sigma \sigma|;\beta ')} + 2 (\rb+\rhob)^2 u_{(\alpha '}h^{\R 1}_{|\sigma u|;\beta ')} \nonumber \\
		& + \rb(\rb+\rhob)(\rb+2\rhob) u_{(\alpha '}h^{\R 1}_{|uu|;\beta ')} + 2 (\rb+\rhob) u_{(\alpha '}h^{\R 1}_{\beta ')\sigma;\sigma} + 2(\rb+\rhob)^2 u_{(\alpha '}h^{\R 1}_{\beta ')u;\sigma} - 2 (\rb + \rhob)^2 u_{(\alpha '}h^{\R 1}_{\beta ')\sigma;u} \nonumber \\
		& - 2 \rb (\rb + \rhob) (\rb + 2 \rhob) u_{(\alpha '}\hdot^{\R 1}_{\beta ')u}\bigr)\Bigr]\biggr] + \order{\lambda}. \label{eq:hSRCov}
\end{align}%
\endgroup%
\end{widetext}%
These need to satisfy
\begin{multline}
	\deltaG^{\mu\nu}[h^{\S\R}] + \deltaG^{\mu\nu}[h^{\S1\abold}] + 2\delta^{2}G^{\mu\nu}[h^{\R1},h^{\S1\acan}] \\
		= \order{\lambda^0}, \quad x\notin \gamma. \label{eq:hSRCovEFE}
\end{multline}
We have successfully checked that the covariant punctures for \(h^{\S\R}_{\mu\nu}\) and \(h^{\S1,\abold}_{\mu\nu}\) satisfy Eq.~\eqref{eq:hSRCovEFE} through the leading two orders, \(\lambda^{-3}\) and \(\lambda^{-2}\).
However, we have not been able to verify this at the highest order we have calculated, order \(\lambda^{-1}\).
This is due to the complexity and length of the expressions when taking multiple different combinations of derivatives.
Despite this, we provide all orders of the covariant punctures for the different singular field terms in a \textsc{Mathematica} notebook in the Supplemental Material~\cite{SuppMat}.

Comparing the covariant puncture for \(h^{\S1}_{\mu\nu}\) from Eq.~\eqref{eq:hS1NoAccCov} to the Lorenz gauge version of the puncture from Eq.~(127) of \citetalias{Pound2014},
\begin{equation}
	h^{\S1\acan,\mathrm{Lor}}_{\alpha\beta} = \frac{2m}{\lambda\rhob}g^{\alpha'}{}_{\alpha}g^{\beta'}{}_{\beta}(g_{\alpha'\beta'} + 2u_{\alpha'}u_{\beta'}) + \order{\lambda}, \label{eq:hS1NoAccLorRov}
\end{equation}
we see that the highly regular gauge puncture has a more complicated form.
This continues at higher order with the Lorenz gauge puncture being substantially simpler and shorter at all orders.
The more complex form results from the highly regular gauge conditions that seek to preserve the background light cone structure emanating from the worldline in the perturbed spacetime; see Sec.~\ref{sec:hr} for further discussion.
This has the knock-on effect that the coordinate expansion in the highly regular gauge will be much more complicated than the Lorenz gauge one as we are introducing more and more terms, and more quantities will need to be expanded.
Thus, if we wanted to perform a mode decomposition of the singular field in the highly regular gauge, we would find that the process is likely to be more complicated than in the Lorenz gauge due to an increase in the number of quantities that need to be decomposed into modes.
However, we believe that the benefits of the highly regular gauge outweigh any disadvantages that may come from the metric perturbations having a more complicated structure.
Merely eliminating the two leading orders of \(h^{\S\S}_{\mu\nu}\) in Eq.~\eqref{eq:hSSCov} has dramatic consequences as it alleviates the problem of infinite mode coupling~\cite{Miller2016} that was discussed in the introduction.
This should allow one to much more efficiently calculate modes of the second-order source.

\section{Coordinate expansion}\label{sec:CoordExp}

In order to implement the covariant expansions in a specific calculation, one must first write them in a chosen coordinate system.
This necessitates re-expanding all the covariant quantities in terms of coordinate differences,
\begin{equation}
	\Dx^{\alpha'} \coloneqq x^{\alpha} - x^{\alpha'}, \label{eq:Dxdef}
\end{equation}
where \(\Dx^{\alpha'} \sim \lambda\).
A derivative of \(\Dx^{\alpha'}\) at \(x^{\mu'}\) then gives
\begin{equation}
	\Dx^{\alpha'}{}_{,\beta'} = -\delta^{\alpha'}_{\beta'}. \label{eq:Dxdiff}
\end{equation}
This leaves us with coefficients evaluated at \(x^{\mu'}\), as in Eq.~\eqref{eq:xxprimeexp}, contracted into certain combinations of \(\Dx^{\alpha'}\).

\subsection{Expanding Synge's world function and the parallel propagator}\label{sec:bitensorExp}

In this section, we generate generic coordinate expansions of the covariant quantities appearing in the punctures from Sec.~\ref{sec:HRCovPuncFinal}.
We begin by expanding Synge's world function, \(\sigma_{\mu'}\), and then use that to find expansions for \(\rb\) and \(\rhob\).
We then move on to find the coordinate expansion for the parallel propagator.

To find a coordinate expansion of Synge's world function, we exploit the fact that it satisfies the identity from Eq.~\eqref{eq:syngeIdentity}.
We make the following ansatz as an expansion for Synge's world function,
\begin{align}
	\sigma ={}& \sum_{n=2}^{\infty} \lambda^{n} A^{(n-1)}_{\alpha'_1\ldots\alpha'_n}(x')\dx^{\alpha'_1}\cdots\dx^{\alpha'_{n}} \nonumber \\
	={}& \lambda^{2}A^{(1)}_{\Delta\Delta}(x') + \lambda^{3}A^{(2)}_{\Delta\Delta\Delta}(x') + \lambda^{4}A^{(3)}_{\Delta\Delta\Delta\Delta}(x') \nonumber \\
		& + \lambda^{5}A^{(4)}_{\Delta\Delta\Delta\Delta\Delta}(x') + \order{\lambda^{6}}; \label{eq:sigmaExp}
\end{align}
see Refs.~\cite{Ottewill2009,Heffernan2012} for similar expansions but with different conventions for \(\Dx^{\alpha'}\).
The primed derivative is then given by
\begin{align}
	\sigma_{\mu'} ={}& -2\lambda A^{(1)}_{\mu'\Delta} + \lambda^{2}\bigl(A^{(1)}_{\Delta\Delta,\mu'} - 3A^{(2)}_{\mu'\Delta\Delta}\bigr) \nonumber \\
		& + \lambda^{3}\bigl(A^{(2)}_{\Delta\Delta\Delta,\mu'} - 4A^{(3)}_{\mu'\Delta\Delta\Delta}\bigr) \nonumber \\
		& + \lambda^{4}\bigl(A^{(3)}_{\Delta\Delta\Delta\Delta,\mu'} - 5A^{(4)}_{\mu'\Delta\Delta\Delta\Delta}\bigr) + \order{\lambda^5}. \label{eq:sigDA}
\end{align}

We then substitute Eqs.~\eqref{eq:sigmaExp}--\eqref{eq:sigDA} into the identity for Synge's world function from Eq.~\eqref{eq:syngeIdentity} and solve order-by-order.
The expressions for \(A^{(n)}_{{\alpha'}^1\cdots{\alpha'}^{n}}\) are
\begin{subequations}
\label{eq:An}
\begin{align}
	A^{(1)}_{\alpha'\beta'} ={}& \frac{1}{2}g_{\alpha'\beta'}, \label{eq:A1} \\
	A^{(2)}_{\alpha'\beta'\gamma'} ={}& \frac{1}{2}g_{\delta'(\alpha'}\Gamma^{\delta'}_{\beta'\gamma')}, \label{eq:A2} \\
	A^{(3)}_{\alpha'\beta'\gamma'\delta'} ={}& \frac{1}{72}\bigl(R_{\alpha'[\gamma'\delta']\beta'} + 3g_{\alpha'\iota'}\Gamma^{\iota'}_{\gamma'\delta',\beta'} \nonumber \\
		& + 9g_{\iota'(\beta'}\Gamma^{\iota'}_{\gamma'\delta'),\alpha'} + 9 g_{\iota'\mu'}\Gamma^{\iota'}_{\alpha'(\beta'}\Gamma^{\mu'}_{\gamma'\delta')} \nonumber \\
		& + 6g_{\mu'(\alpha'}\Gamma^{\mu'}_{\beta')\iota'}\Gamma^{\iota'}_{\gamma'\delta'} + 6g_{\mu'(\gamma'}\Gamma^{\iota'}_{\delta')\beta'}\Gamma^{\mu'}_{\alpha'\iota'}\bigr), \label{eq:A3}  \displaybreak[0] \\
	A^{(4)}_{\alpha'\beta'\gamma'\delta'\iota'} ={}& \frac{1}{120} \bigl(5 g_{(\alpha ' |\rho ' |}\Gamma^{\rho '}_{\delta ' \iota ',\gamma'\beta')} \nonumber \\
		& + 5 \Gamma^{\rho '}_{(\alpha ' \beta '}g_{\gamma ' |\kappa ' |}\Gamma^{\kappa '}_{\iota '|\rho '|,\delta')} \nonumber \\
		& + 10 \Gamma^{\rho '}_{(\alpha ' \beta '}g_{|\rho ' \kappa ' |}\Gamma^{\kappa '}_{\delta ' \iota ',\gamma')} \nonumber \\
		& + 10 \Gamma^{\rho '}_{(\alpha ' |\kappa ' |}g_{\beta ' |\rho ' |}\partial_{\gamma '}\Gamma^{\kappa '}_{\delta ' \iota ')} \nonumber \\
		& + 3 \Gamma^{\rho '}_{(\alpha ' \beta '}\Gamma^{\kappa '}_{\gamma ' \delta '}\Gamma^{\mu '}_{\iota ')\rho '}g_{\kappa ' \mu '} \nonumber \\
		& + 7 \Gamma^{\rho '}_{(\alpha ' \beta '}\Gamma^{\kappa '}_{\gamma ' \delta '}\Gamma^{\mu '}_{\iota ')\kappa '}g_{\rho ' \mu '} \nonumber \\
		& + 5 \Gamma^{\rho '}_{(\alpha ' \beta '}\Gamma^{\kappa '}_{\gamma ' |\mu ' |}\Gamma^{\mu '}_{\delta ' |\rho ' |}g_{\iota ')\kappa '}\bigr). \label{eq:A4}
\end{align}%
\end{subequations}
These are similar to the expansions appearing in Eq.~(2.10) of Ref.~\cite{Ottewill2009} and Eq.~(3.10) of Ref.~\cite{Heffernan2012}, but here, we have a slightly different definition for \(\Dx^{\alpha'}\) and we take the derivatives at \(x^{\mu'}\) instead of \(x^\mu\).
Taking the primed derivative of the appropriate quantities and then substituting these and Eq.~\eqref{eq:An} into Eq.~\eqref{eq:sigDA} gives us the final expression for the coordinate expansion of Synge's world function,
\begin{equation}
	\sigma_{\alpha'} = \sum_{n=1}^{\infty}\lambda^{n}\sigma^{(n)}_{\alpha'}, \label{eq:sigDSum}
\end{equation}
where the first four orders are given by
\begin{subequations}
	\label{eq:sign}
	\begin{align}
		\sigma^{(1)}_{\alpha'} ={}& -\dx_{\alpha'}, \label{eq:sig1} \\
		\sigma^{(2)}_{\alpha'} ={}& -\frac{1}{2}g_{\alpha'\delta'}\Gamma^{\delta'}_{\Delta\Delta}, \label{eq:sig2} \\
		\sigma^{(3)}_{\alpha'} ={}& -\frac{1}{6}\bigl(g_{\alpha'\iota'}\Gamma^{\iota'}_{\Delta\Delta,\Delta} + g_{\alpha'\mu'}\Gamma^{\iota'}_{\Delta\Delta}\Gamma^{\mu'}_{\Delta\iota'}\bigr), \label{eq:sig3} \displaybreak[0] \\
		\sigma^{(4)}_{\alpha'} ={}& -\frac{1}{24}\bigl[\Gamma^{\nu'}_{\Delta\Delta}\bigl(g_{\alpha'\mu'}\Gamma^{\kappa'}_{\Delta\nu'}\Gamma^{\mu'}_{\Delta\kappa'} + g_{\alpha'\kappa'}\Gamma^{\kappa'}_{\Delta\nu',\Delta} \nonumber \\
			& - R_{\alpha'\Delta\Delta\nu'}\bigr) + g_{\alpha'\nu'}\bigl(2\Gamma^{\nu'}_{\Delta\kappa'}\Gamma^{\kappa'}_{\Delta\Delta,\Delta} + \Gamma^{\nu'}_{\Delta\Delta,\Delta\Delta}\bigr)\bigr]. \label{eq:sig4}
	\end{align}%
\end{subequations}
To check these expressions, one can substitute Eq.~\eqref{eq:sign} into Eq.~\eqref{eq:syngeIdentity} to demonstrate they satisfy the identity for Synge's world function.

We also require the expansions of \(\rb\) and \(\rhob\) from Eqs.~\eqref{eq:rdef}--\eqref{eq:rhodef} which can be performed by substituting in Eqs.~\eqref{eq:sigDSum}--\eqref{eq:sign}.
The expression for \(\rb\) is trivial as it just requires us to contract the four-velocity into Eq.~\eqref{eq:sigDSum}, so that, at leading order,
\begin{equation}
	\rb = -\lambda\rbo + \order{\lambda^2}, \label{eq:rdefr0}
\end{equation}
where, in analogy with Eq.~\eqref{eq:rdef}, we define the four-velocity contracted with the coordinate difference as
\begin{equation}
	\rbo \coloneqq u_{\mu'}\dx^{\mu'}, \label{eq:r0def}
\end{equation}

We write the expansion of \(\rhob\) as a power series,
\begin{equation}
	\rhob = \sum_{n=1}^{\infty} \lambda^{n}\rhob^{(n)}, \label{eq:rhoSum}
\end{equation}
and define
\begin{equation}
	\rhob_0 \coloneqq \sqrt{P_{\mu'\nu'}\dx^{\mu'}\dx^{\nu'}}. \label{eq:rho0def}
\end{equation}
We then proceed to substitute our coordinate expansion for \(\sigma_{\alpha'}\) from Eq.~\eqref{eq:sigDSum} into the definition for \(\rhob\) from Eq.~\eqref{eq:rhoSum} and collect terms at each order in \(\lambda\).
The first four orders of the expansion are given by
\begin{subequations}
\label{eq:rhon}
\begin{align}
	\rhob^{(1)} ={}& \rhobo, \label{eq:rho1} \\
	\rhob^{(2)} ={}& \frac{1}{2\rhobo}\bigl(\chrxxx + \chruxx \rbo\bigr), \label{eq:rho2} \\
	\rhob^{(3)} ={}& -\frac{1}{8\rhobo^3}\bigl(\chrxxx+\chruxx\rbo\bigr)^2 + \frac{1}{24\rhobo}\bigl(3\chruxx^2 \nonumber \\
		& + 4\chrxxxx + 4\chruxxx\rbo + 4\rbo\chrBxx{\alpha'}\chruBx{\alpha'} \nonumber \\
		& + 4\chrBxx{\alpha'}\chrxBx{\alpha'} + 3g_{\alpha'\beta'}\chrBxx{\alpha'}\chrBxx{\beta'}\bigr), \label{eq:rho3} \displaybreak[0] \\
	\rhob^{(4)} ={}& \frac{1}{16\rhobo^5}\bigl(\chrxxx+\chruxx\rbo\bigr)^3 - \frac{1}{48\rhobo^3}\bigl(\Gamma_{\Delta \Delta}^{\Delta} + \Gamma_{\Delta \Delta}^{u} \rbo\bigl) \nonumber \\
		& \times \bigl(3 {\Gamma_{\Delta \Delta}^{u}}^2 + \Gamma_{\Delta \Delta}^{\alpha'} \bigl[4 \Gamma_{\alpha'\Delta}^{\Delta} + 3 g_{\alpha'\beta'}\Gamma_{\Delta \Delta}^{\beta'} \nonumber \\
		& + 4 \Gamma_{\alpha'\Delta}^{u} \rbo\bigr] + 4 \bigl[\Gamma_{\Delta \Delta , \Delta}^{\Delta} + \Gamma_{\Delta \Delta , \Delta}^{u} \rbo\bigr]\bigr) \nonumber \\
		& + \frac{1}{24\rhobo}\bigl(2 \Gamma_{\Delta \Delta}^{u} \Gamma_{\Delta \Delta , \Delta}^{u} + 2 \Gamma_{\Delta \Delta , \Delta}^{\alpha'} \Gamma_{\alpha'\Delta}^{\Delta} + \Gamma_{\Delta \Delta , \Delta \Delta}^{\Delta} \nonumber \\
		& + 2 \Gamma_{\Delta \Delta , \Delta}^{\alpha'} \Gamma_{\alpha'\Delta}^{u} \rbo + \Gamma_{\Delta \Delta , \Delta \Delta}^{u} \rbo \nonumber \\
		& + \Gamma_{\Delta \Delta}^{\alpha'} \bigl[2 \Gamma_{\alpha'\Delta}^{u} \Gamma_{\Delta \Delta}^{u} + \Gamma_{\Delta \alpha', \Delta}^{\Delta} + 2  g_{\alpha'\beta'}\Gamma_{\Delta \Delta , \Delta}^{\beta'} \nonumber \\
		& + \Gamma_{\Delta \alpha', \Delta}^{u} \rbo + R_{\alpha'\Delta u\Delta} \rbo\bigr] + \Gamma_{\Delta\beta'}^{\alpha'} \Gamma_{\Delta \Delta}^{\beta'} \bigl[\Gamma_{\alpha'\Delta}^{\Delta} \nonumber \\
		& + 2 g_{\alpha'\gamma'}\Gamma_{\Delta \Delta}^{\gamma'} + \Gamma_{\alpha'\Delta}^{u} \rbo\bigr]\bigr). \label{eq:rho4}
\end{align}%
\end{subequations}

To calculate the coordinate expansion of \(g^{\nu'}{}_{\mu}\), we proceed in a similar way to that of \(\sigma_{\alpha'}\).
To begin, we use the ansatz
\begin{multline}
	g^{\nu'}{}_{\mu} = \delta^{\nu'}_{\mu'} + \lambda G^{(1)\nu'}{}_{\mu'\Delta} + \lambda^{2} G^{(2)\nu'}{}_{\mu'\Delta\Delta} \\
	+ \lambda^{3} G^{(3)\nu'}{}_{\mu'\Delta\Delta\Delta} + \order{\lambda^4} \label{eq:ppAnsatz}
\end{multline}
and substitute this into the identity for the derivative of the parallel propagator contracted into a derivative of Synge's world function from Eq.~\eqref{eq:SigPPDContract}.
We proceed to solve this order-by-order to find
\begin{subequations}
\label{eq:Gn}
\begin{align}
	G^{(1)\alpha'}{}_{\beta'\gamma'} ={}& \Gamma^{\alpha'}_{\beta'\gamma'}, \label{eq:G1} \\
	G^{(2)\alpha'}{}_{\beta'\gamma'\delta'} ={}& \frac{1}{2}\bigl(\Gamma^{\alpha'}_{\beta'\iota'}\Gamma^{\iota'}_{\gamma'\delta'} + R^{\alpha'}{}_{(\gamma'\delta')\beta'} + \Gamma^{\alpha'}_{\gamma'\delta',\beta'}\bigr), \label{eq:G2} \\
	G^{(3)\alpha'}{}_{\beta'\gamma'\delta'\iota'} ={}& \frac{1}{6}\sym_{\gamma'\delta'\iota'}\bigl(\Gamma^{\varsigma'}_{\gamma'\delta'}\bigl[3R^{\alpha'}{}_{(\varsigma'\iota')\beta'} + \Gamma^{\alpha'}_{\iota'\varsigma',\beta'}\bigr] \nonumber \\
		& + \Gamma^{\alpha'}_{\beta'\varsigma'}\bigl[\Gamma^{\varsigma'}_{\gamma'\kappa'}\Gamma^{\kappa'}_{\delta'\iota'} + \Gamma^{\varsigma'}_{\gamma'\delta',\iota'}\bigr] \nonumber \\
		& - \Gamma^{\varsigma'}_{\beta'\gamma'}R^{\alpha'}{}_{\iota'\varsigma'\delta'} + R^{\alpha'}{}_{\delta'\gamma'\beta';\iota'} \nonumber \\
		& + \Gamma^{\alpha'}_{\gamma'\varsigma'}\Gamma^{\varsigma'}_{\delta'\iota',\beta'} + \Gamma^{\alpha'}_{\gamma'\delta',\beta'\iota'}\bigr). \label{eq:G3}
\end{align}%
\end{subequations}
As with Eq.~\eqref{eq:An}, similar expansions of the parallel propagator have been done previously in Eqs.~(3.10)--(3.12) of Ref.~\cite{Heffernan2022}.
We have checked our expressions by substituting them into Eq.~\eqref{eq:SigPPDContract} and have verified that they satisfy the identity to the appropriate order in \(\lambda\).

\subsection{Coordinate expansions of the covariant punctures}\label{sec:CoordPunc}

With our covariant punctures derived, we can proceed to write them as a generic coordinate expansion using the techniques discussed for the singular scalar field in Sec.~\ref{sec:bitensorExp}.
This will allow them to be easily written in any desired coordinate system.

To do so, we substitute our coordinate expansion for \(\sigma_{\alpha'}\) from Eqs.~\eqref{eq:sigDSum}--\eqref{eq:sign}, \(\rhob\) from Eqs.~\eqref{eq:rhoSum}--\eqref{eq:rhon} and \(g^{\mu'}{}_{\mu}\) from Eqs.~\eqref{eq:ppAnsatz}--\eqref{eq:Gn} into the expression for \(h^{\S}_{\mu\nu}\) from Sec.~\ref{sec:HRCovPuncFinal}.
Doing so results in expressions that are written in terms of the coordinate difference, \(\Delta x^{\mu'}\) and the four-velocity, \(u^{\mu'}\) along with \(h^{\R1}_{\mu'\nu'}\), \(\Gamma^{\mu'}_{\nu'\rho'}\), and \(R_{\alpha'\beta'\mu'\nu'}\) and their respective derivatives.
The final expressions are incredibly long and, as such, we only display them through order \(\lambda^0\) (except for \(h^{\S\R}_{\mu\nu}\), for which we just display the leading-order term).
The higher order terms are available in the Supplemental Material in a \textsc{Mathematica} notebook~\cite{SuppMat}.

The coordinate expansion of the first-order singular field, with no acceleration, in the highly regular gauge is given by
\begin{align}
	h^{\S1\acan}_{\mu\nu} ={}& \frac{2 m}{\lambda\rhobo^3} \bigl(\Delta x_{\mu '} + u_{\mu '} (\rbo -  \rhobo)\bigr) \bigl(\Delta x_{\nu '} + u_{\nu '} (\rbo -  \rhobo)\bigr) \nonumber \\
		& - \frac{m\lambda^0}{\rhobo^5}\bigl[u_{\mu '} u_{\nu '} (\rbo -  \rhobo) \bigl(3 \rbo (\Gamma_{\Delta \Delta}^{\Delta} + \Gamma_{\Delta \Delta}^{u} \rbo) \nonumber \\
		& -  (\Gamma_{\Delta \Delta}^{\Delta} + \Gamma_{\Delta \Delta}^{u} \rbo) \rhobo - 2 \Gamma_{\Delta \Delta}^{u} \rhobo^2\bigr) + 3 \Delta x_{\mu '} \Delta x_{\nu '} \nonumber \\
		& \times (\Gamma_{\Delta \Delta}^{\Delta} + \Gamma_{\Delta \Delta}^{u} \rbo) + 2\bigl(\Gamma_{\Delta \Delta}^{\Delta} (3 \rbo - 2 \rhobo) \nonumber \\
		& + \Gamma_{\Delta \Delta}^{u} (\rbo -  \rhobo) (3 \rbo + \rhobo)\bigr)u_{(\mu '}\Delta x_{\nu ')} \nonumber \\
		& - 2 u_{(\mu '} \bigl(2 \Gamma^{\Delta }_{\nu ') \Delta } + g_{\nu ') \alpha '}\Gamma^{\alpha '}_{\Delta \Delta }  + 2 \Gamma^{u}_{\nu ') \Delta } (\rbo -  \rhobo)\bigr) \nonumber \\
		& \times (\rbo -  \rhobo) \rhobo^2 - 2 \Delta x_{(\mu '} \rhobo^2 \bigl(2 \Gamma^{\Delta }_{\nu ') \Delta } +  g_{\nu ') \alpha '}\Gamma^{\alpha '}_{\Delta \Delta } \nonumber \\
		& + 2 \Gamma^{u}_{\nu ') \Delta } (\rbo - \rhobo)\bigr)\bigr] + \order{\lambda}. \label{eq:hS1NoAccCoordExp}
\end{align}

Moving to second order, the first-order singular field with acceleration is
\begin{align}
	h^{\S1\abold}_{\mu\nu} ={}& - \frac{m\lambda^0}{\rhobo^5}\bigl[a_{\Delta}{} u_{\mu '} u_{\nu '} (3 \rbo^3 - \rbo^2 \rhobo - 4 \rbo \rhobo^2 + 4 \rhobo^3) \nonumber \\
		& (\rhobo-\rbo) - 3 \Delta x_{\mu '} \Delta x_{\nu '} a_{\Delta}{} \rbo^2 + 2 \bigl(\rbo (\rbo - 2 \rhobo) \nonumber \\
		& \times \rhobo^2 \Delta x_{(\mu '}a_{\nu ')} + a_{\Delta}{} (2 \rbo^2 \rhobo + 2 \rbo \rhobo^2 \nonumber \\
		& - 2 \rhobo^3 - 3 \rbo^3) \Delta x_{(\mu '}u_{\nu ')} + \rbo (\rbo - 2 \rhobo) \nonumber \\
		& (\rbo -  \rhobo) \rhobo^2 a_{(\mu '}u_{\nu ')}\bigr)\bigr] + \order{\lambda}. \label{eq:hS1AccCoordExp}
\end{align}
The ``singular times singular'' piece is given by
\begin{align}
	h^{\S\S}_{\mu\nu} ={}& - \frac{2 m^2\lambda^0}{3 \rhobo^4} \bigl[6 \rhobo^3 R_{(\mu ' |\Delta |\nu ')u} + 2 (3 \rbo -  \rhobo) \rhobo^3 \nonumber \\
		& \times R_{(\mu ' |u|\nu ')u} + R_{\Delta u\Delta u} \bigl(5 \Delta x_{\mu '} \Delta x_{\nu '} -  g_{\mu ' \nu '} \rhobo^2 \nonumber \\
		& + u_{\mu '} u_{\nu '} (5 \rbo^2 - 6 \rbo \rhobo + 5 \rhobo^2) + (10 \rbo - 6 \rhobo) \nonumber \\
		& \times \Delta x_{(\mu '}u_{\nu ')}\bigr) - 6 \rhobo \Delta x_{(\mu '}R_{\nu ')\Delta \Delta u} + 2\rhobo \nonumber \\
		& \times (2 \rhobo - 3 \rbo) \Delta x_{(\mu '}R_{\nu ')u\Delta u} - 6 \rbo \rhobo u_{(\mu '}R_{\nu ')\Delta \Delta u} \nonumber \\
		& - 2 \rhobo (3 \rbo^2 - 2 \rbo \rhobo + 3 \rhobo^2) u_{(\mu '}R_{\nu ')u\Delta u}\bigr] + \order{\lambda}. \label{eq:hSSCoordExp}
\end{align}
Finally, the ``singular times regular'' piece is
\begin{align}
	h^{\S\R}_{\mu\nu} ={}& \frac{m}{\lambda\rhobo^5}\bigl[4 \rhobo^2 \Delta x_{(\mu '}h^{\R 1}_{\nu ')\Delta } + 4 (\rbo -  \rhobo) \rhobo^2 \Delta x_{(\mu '}h^{\R 1}_{\nu ')u} \nonumber \\
		& - h^{\R 1}_{\Delta u} \bigl(6 \Delta x_{\mu '} \Delta x_{\nu '} \rbo + 2 u_{\mu '} u_{\nu '} (\rbo -  \rhobo)^2 \nonumber \\
		& \times (3 \rbo + 2 \rhobo) + 4 (\rbo -  \rhobo) (3 \rbo + \rhobo) \Delta x_{(\mu '}u_{\nu ')}\bigr) \nonumber \\
		& - h^{\R 1}_{\Delta \Delta } \bigl(3 \Delta x_{\mu '} \Delta x_{\nu '} +  u_{\mu '} u_{\nu '} (\rbo -  \rhobo) (3 \rbo -  \rhobo) \nonumber \\
		& + 2(3 \rbo - 2 \rhobo) \Delta x_{(\mu '}u_{\nu ')}\bigr) - h^{\R 1}_{uu} \bigl(3 \Delta x_{\mu '} \Delta x_{\nu '} \rbo^2 \nonumber \\
		& + u_{\mu '} u_{\nu '} (\rbo -  \rhobo)^2 (3 \rbo^2 + 2 \rbo \rhobo - 2 \rhobo^2) \nonumber \\
		& + 2 (\rbo -  \rhobo) (3 \rbo^2 + \rbo \rhobo -  \rhobo^2) \Delta x_{(\mu '}u_{\nu ')}\bigr) \nonumber \\
		& + 4 (\rbo -  \rhobo) \rhobo^2 u_{(\mu '}(h^{\R 1}_{\nu ')\Delta} + (\rbo -  \rhobo)h^{\R 1}_{\nu ')u})\bigr] \nonumber \\
		& + \order{\lambda^0}. \label{eq:hSRCoordExp}
\end{align}

\section{Conclusion and applications}\label{sec:conclusion}

The main result of this paper is the conversion of the local coordinate form of the metric perturbations given in \citetalias{Upton2021} into fully covariant form using the methods of \citetalias{Pound2014}.
These were provided in truncated form in Sec.~\ref{sec:HRCovPuncFinal} and in full form in the \textsc{Mathematica} notebook in the Supplemental Material~\cite{SuppMat}.

We have then re-expanded these covariant expressions and written them as a generic coordinate expansion that is valid in any desired coordinate system.
As with the covariant expressions, abridged forms were presented in Sec.~\ref{sec:CoordPunc}, with the full expressions appearing in the Supplemental Material~\cite{SuppMat}.
By providing the metric perturbations in these forms, we have enabled them to be written in any desired coordinate system without necessitating the use of a potentially complicated coordinate transformation from Fermi--Walker coordinates.

One useful immediate extension of this work would be to calculate the modes of the punctures to see how well the highly regular gauge alleviates the problem of infinite mode coupling.
For quasicircular orbits in Schwarzschild, for example, one could decompose the punctures into modes using the methods of Ref.~\cite{WardellWarburton2015}.
From this, one could use the mode coupling formula from Eq.~\eqref{eq:d2Gilm} to explicitly calculate the behaviour of the second-order Einstein tensor near to the worldline of the small object.

An interesting property of the highly regular gauge to note is that, following from the gauge conditions given in Sec.~\ref{sec:hr}, one can write the singular field metric perturbations in terms of null vectors.
For example, if one defines
\begin{align}
    k_{\alpha} ={}& \frac{g^{\alpha'}{}_{\alpha}}{\sqrt{2}}\biggl(u_{\alpha'} + \frac{P_{\alpha'\beta'}\sigma^{\beta'}}{\rhob}\biggr) \nonumber \\
    ={}& \frac{g^{\alpha'}{}_{\alpha}}{\sqrt{2}\rhob}\bigl(\sigma_{\alpha'}+(\rb+\rhob)u_{\alpha'}\bigr), \label{eq:kNullDef}
\end{align}
so that \(k^{\alpha}k_{\alpha}=0\), one can write the first-order singular field from Eq.~\eqref{eq:hS1NoAccCov} as
\begin{equation}
    h^{\S1\acan}_{\mu\nu} = \frac{4m}{\lambda\rhob}k_{\mu}k_{\nu} + \order{\lambda}. \label{eq:hS1KS}
\end{equation}
One can then write Eq.~\eqref{eq:fullg} in terms of these null vectors as
\begin{align}
    \fullg_{\mu\nu} ={}& g_{\mu\nu} + \e (h^{\R1}_{\mu\nu} + h^{\S1\acan}_{\mu\nu}) + \order{\e^2}, \nonumber \\
    ={}& g_{\mu\nu} + \e \lambda^{-1} 2Vk_{\mu}k_{\nu} + \order{\e\lambda^0,\e^2}, \label{eq:fullgApproxKS}
\end{align}
where \(V=2m/(\rhob)\).
This has the form of a Kerr--Schild perturbation~\cite{Kerr1965,*[Reproduced in: ]Kerr2009,Taub1981} on the background spacetime.
However, this correspondence is broken in the singular field at order \(\lambda\) through the introduction of Riemann tidal terms in \(h^{\S1}_{\mu\nu}\).
Additionally, \(h^{\R1}_{\mu\nu}k^{\mu}k^{\nu}\neq 0\) due to the regular field being in a generic gauge.

It would be interesting to further explore the connection between the highly regular gauge and Kerr--Schild gauges, potentially drawing on previous work by \textcite{Harte2014} and \textcite{Harte2016}.

\begin{acknowledgments}
We thank Adam Pound for numerous helpful discussions and for providing the \textsc{Mathematica}/\textsc{xAct} code that was used in \citetalias{Pound2014}.

This work was supported by a Royal Society Research Grant for Research Fellows and the fellowship Lumina Quaeruntur No. LQ100032102 of the Czech Academy of Sciences.
\end{acknowledgments}

\appendix

\section{Quasi-Kerr--Schild form of the metric perturbations}

In this section, we derive the leading-order form of the second-order metric perturbations when written in the quasi-Kerr--Schild form discussed in Sec.~\ref{sec:conclusion}.

We begin with the first-order singular field with acceleration.
This now takes the form
\begin{align}
	h^{\S1\abold}_{\mu\nu} ={}& \frac{\lambda^0 V}{\sqrt{2}\rhob^2}\bigl[2a_{(\mu}k_{\nu)}\rb\rhob(\rb+2\rhob) \nonumber \\
		& + \bigl\{a^k(\rb-\rhob)+a^N(\rb+\rhob)\bigr\} \nonumber \\
		& \times \bigl\{2k_{\mu}k_{\nu}(\rb^2-\rb\rhob-\rhob^2) \nonumber \\
		& - k_{(\mu}N_{\nu)}(\rb^2+2\rb\rhob+2\rhob^2)\bigr\}\bigr] + \order{\lambda},
\end{align}
where we have contracted in the parallel propagators and introduced the auxiliary null vector
\begin{align}
	N_\alpha ={}& \frac{g^{\alpha'}{}_{\alpha}}{\sqrt{2}}\biggl(u_{\alpha'} - \frac{P_{\alpha'\beta'}\sigma^{\beta'}}{\rhob}\biggr) \nonumber \\
	={}& - \frac{g^{\alpha'}{}_{\alpha}}{\sqrt{2}\rhob}\bigl(\sigma_{\alpha'}+(\rb-\rhob)u_{\alpha'}\bigr) \label{eq:NNullDef}
\end{align}
that is normalised so that \(k^\alpha N_{\alpha} = -1\).
Moving on to the ``singular times singular'' piece, after substituting in Eqs.~\eqref{eq:kNullDef} and~\eqref{eq:NNullDef}, we see that
\begin{align}
	h^{\S\S}_{\mu\nu} ={}& \frac{\lambda^0\rhob^2V^2}{6}\bigl[4R_{\mu k\nu k} - 2R_{k (\mu\nu) N} - 2R_{\mu N\nu N} \nonumber \\
	& + 2(4k_{(\mu}-N_{(\mu})R_{\nu)kNk} - 2(k_{(\mu} - 2N_{(\mu})R_{\nu NkN} \nonumber \\
	& + (g_{\mu\nu} - 8k_{\mu}k_{\nu} - 2N_{\mu}N_{\nu})R_{kNkN}\bigr] + \order{\lambda}. \label{eq:hSSKS}
\end{align}
Finally, the ``singular times regular'' piece is now given by
\begin{align}
	h^{\S\R}_{\mu\nu} ={}& \frac{V}{2\lambda}\bigl[8k_{(\mu}h^{\R1}_{\nu)k} + \bigl(h^{\R1}_{kk}+6h^{\R1}_{kN}-3h^{\R1}_{NN}\bigr)k_{\mu}k_{\nu} \nonumber \\
		& + 4h^{\R1}_{kk}k_{(\mu}N_{\nu)}\bigr] + \order{\lambda^0}. \label{eq:hSRKS}
\end{align}
These calculations can be extended to higher order in \(\lambda\) by continuing to replace \(u_{\alpha'}\) and \(\sigma_{\alpha'}\) with \(k_{\alpha}\) and \(N_{\alpha}\) through the use of Eqs.~\eqref{eq:kNullDef} and~\eqref{eq:NNullDef}.

\bibliography{bibfile}

\end{document}